\documentclass[twocolumn,twocolappendix,trackchanges]{aastex701}

\usepackage{newtxtext,newtxmath}
\usepackage[T1]{fontenc}

\newcommand{\Msunh}{\>h^{-1} {\rm M_\odot}}

\newcommand{\mpch}{\>h^{-1}{\rm {Mpc}}}

\newcommand{\kms}{\>{\rm km}\,{\rm s}^{-1}}

\usepackage{graphicx}	% Including figure files
\usepackage{amsmath}	% Advanced maths commands

\usepackage{color}
\usepackage{lastpage}
\usepackage{bm}
\usepackage{placeins}

\begin{document}

\title{ELUCID-DESI II. Revealing dark matter mass, tidal, and velocity (MTV) fields  using galaxy group phase information}

\author[orcid=0000-0003-0771-1350]{Qingyang Li}
% \altaffiliation{Kitt Peak National Observatory}
\affiliation{State Key Laboratory of Dark Matter Physics, Tsung-Dao Lee Institute \& School of Physics and Astronomy, Shanghai Jiao Tong University, Shanghai 201210, China}
\affiliation{Shanghai Key Laboratory for Particle Physics and Cosmology, and Key Laboratory for Particle Physics, Astrophysics and Cosmology, Ministry of Education, Shanghai Jiao Tong University, Shanghai 200240, China}
\email[show]{qingyli@sjtu.edu.cn}  

\author[orcid=0000-0003-3997-4606]{Xiaohu Yang}
\affiliation{State Key Laboratory of Dark Matter Physics, Tsung-Dao Lee Institute \& School of Physics and Astronomy, Shanghai Jiao Tong University, Shanghai 201210, China}
\affiliation{Shanghai Key Laboratory for Particle Physics and Cosmology, and Key Laboratory for Particle Physics, Astrophysics and Cosmology, Ministry of Education, Shanghai Jiao Tong University, Shanghai 200240, China}
\email[show]{xyang@sjtu.edu.cn} 

\author{Wensheng Hong}
\affiliation{State Key Laboratory of Dark Matter Physics, Tsung-Dao Lee Institute \& School of Physics and Astronomy, Shanghai Jiao Tong University, Shanghai 201210, China}
\affiliation{Shanghai Key Laboratory for Particle Physics and Cosmology, and Key Laboratory for Particle Physics, Astrophysics and Cosmology, Ministry of Education, Shanghai Jiao Tong University, Shanghai 200240, China}
\email[]{}

\author[]{Feng Shi}
\affiliation{School of Aerospace Science And Technology, Xidian University, Xi'an 710126, China}
\email[]{}

\author[]{Youcai Zhang}
\affiliation{Shanghai Astronomical Observatory, Nandan Road 80, Shanghai 200030, China}
\email[]{}

\author[]{Jiaqi Wang}
\affiliation{State Key Laboratory of Dark Matter Physics, Tsung-Dao Lee Institute \& School of Physics and Astronomy, Shanghai Jiao Tong University, Shanghai 201210, China}
\affiliation{Shanghai Key Laboratory for Particle Physics and Cosmology, and Key Laboratory for Particle Physics, Astrophysics and Cosmology, Ministry of Education, Shanghai Jiao Tong University, Shanghai 200240, China}
\email[]{}

\author[]{Junde Li}
\affiliation{State Key Laboratory of Dark Matter Physics, Tsung-Dao Lee Institute \& School of Physics and Astronomy, Shanghai Jiao Tong University, Shanghai 201210, China}
\affiliation{Shanghai Key Laboratory for Particle Physics and Cosmology, and Key Laboratory for Particle Physics, Astrophysics and Cosmology, Ministry of Education, Shanghai Jiao Tong University, Shanghai 200240, China}
\email[]{}

\author[]{Yiyang Guo}
\affiliation{State Key Laboratory of Dark Matter Physics, Tsung-Dao Lee Institute \& School of Physics and Astronomy, Shanghai Jiao Tong University, Shanghai 201210, China}
\affiliation{Shanghai Key Laboratory for Particle Physics and Cosmology, and Key Laboratory for Particle Physics, Astrophysics and Cosmology, Ministry of Education, Shanghai Jiao Tong University, Shanghai 200240, China}
\email[]{}

\author{Yingxiao Song}
\affiliation{State Key Laboratory of Dark Matter Physics, Tsung-Dao Lee Institute \& School of Physics and Astronomy, Shanghai Jiao Tong University, Shanghai 201210, China}
\affiliation{Shanghai Key Laboratory for Particle Physics and Cosmology, and Key Laboratory for Particle Physics, Astrophysics and Cosmology, Ministry of Education, Shanghai Jiao Tong University, Shanghai 200240, China}
\email[]{}

\author[]{Huiyuan Wang}
\affiliation{Key Laboratory for Research in Galaxies and Cosmology, Department of Astronomy, University of Science and Technology of China, Hefei, Anhui 230026, China}
\email[]{}

\author[]{Yan-Chuan Cai}
\affiliation{Institute for Astronomy, University of Edinburgh, Royal Observatory, Blackford Hill, Edinburgh, EH9 3HJ, UK}
\email[]{}

\author[]{Yizhou Gu}
\affiliation{State Key Laboratory of Dark Matter Physics, Tsung-Dao Lee Institute \& School of Physics and Astronomy, Shanghai Jiao Tong University, Shanghai 201210, China}
\affiliation{Shanghai Key Laboratory for Particle Physics and Cosmology, and Key Laboratory for Particle Physics, Astrophysics and Cosmology, Ministry of Education, Shanghai Jiao Tong University, Shanghai 200240, China}
\email[]{}

\author[]{Chengze Liu}
\affiliation{State Key Laboratory of Dark Matter Physics, Tsung-Dao Lee Institute \& School of Physics and Astronomy, Shanghai Jiao Tong University, Shanghai 201210, China}
\affiliation{Shanghai Key Laboratory for Particle Physics and Cosmology, and Key Laboratory for Particle Physics, Astrophysics and Cosmology, Ministry of Education, Shanghai Jiao Tong University, Shanghai 200240, China}
\email[]{}

\author[]{Jiaxin Han}
\affiliation{State Key Laboratory of Dark Matter Physics, Tsung-Dao Lee Institute \& School of Physics and Astronomy, Shanghai Jiao Tong University, Shanghai 201210, China}
\affiliation{Shanghai Key Laboratory for Particle Physics and Cosmology, and Key Laboratory for Particle Physics, Astrophysics and Cosmology, Ministry of Education, Shanghai Jiao Tong University, Shanghai 200240, China}
\email[]{}

\author[]{Zhongxu Zhai}
\affiliation{State Key Laboratory of Dark Matter Physics, Tsung-Dao Lee Institute \& School of Physics and Astronomy, Shanghai Jiao Tong University, Shanghai 201210, China}
\affiliation{Shanghai Key Laboratory for Particle Physics and Cosmology, and Key Laboratory for Particle Physics, Astrophysics and Cosmology, Ministry of Education, Shanghai Jiao Tong University, Shanghai 200240, China}
\email[]{}

\author[]{Yu Yu}
\affiliation{State Key Laboratory of Dark Matter Physics, Tsung-Dao Lee Institute \& School of Physics and Astronomy, Shanghai Jiao Tong University, Shanghai 201210, China}
\affiliation{Shanghai Key Laboratory for Particle Physics and Cosmology, and Key Laboratory for Particle Physics, Astrophysics and Cosmology, Ministry of Education, Shanghai Jiao Tong University, Shanghai 200240, China}
\email[]{}

\author[]{Yipeng Jing}
\affiliation{State Key Laboratory of Dark Matter Physics, Tsung-Dao Lee Institute \& School of Physics and Astronomy, Shanghai Jiao Tong University, Shanghai 201210, China}
\affiliation{Shanghai Key Laboratory for Particle Physics and Cosmology, and Key Laboratory for Particle Physics, Astrophysics and Cosmology, Ministry of Education, Shanghai Jiao Tong University, Shanghai 200240, China}
\email[]{}

\author[]{Houjun Mo}
\affiliation{Department of Astronomy, University of Massachusetts Amherst, MA 01003, USA}
\email[]{}

\author[]{Yuyu Wang}
\affiliation{Department of Physics \& Astronomy, Ohio University, Athens, OH 45701, USA}
\email[]{}

\author[]{Hao-Ran Yu}
\affiliation{Department of Astronomy, Xiamen University, Xiamen, Fujian 361005, China}
\email[]{}

\author[]{Yingjie Peng}
\affiliation{Department of Astronomy, School of Physics, Peking University, Beijing 100871, China}
\affiliation{Kavli Institute for Astronomy and Astrophysics, Peking University, Beijing 100871, China}
\email[]{}

\author[]{Weiguang Cui}
\affiliation{Institute for Astronomy, Royal Observatory, Edinburgh EH9 3HJ, UK}
\affiliation{Centro de Investigación Avanzada en Física Fundamental (CIAFF), Facultad de Ciencias, Universidad Autónoma de Madrid, E-28049, Madrid, Spain}
\affiliation{Departamento de F\'{i}sica Te\'{o}rica, Universidad Aut\'{o}noma de Madrid, M\'{o}dulo 15, E-28049 Madrid, Spain}
\email[]{}

\author[]{Qi Guo}
\affiliation{Institute for Frontiers in Astronomy and Astrophysics, Beijing Normal University, Beijing 102206, China}
\affiliation{School of Physics and Astronomy, Beijing Normal University, Beijing 100875, China}
\affiliation{Key Laboratory for Computational Astrophysics, National Astronomical Observatories, Chinese Academy of Sciences, Beijing 100101, China}
\email[]{}

\author[]{Liang Gao}
\affiliation{Institute for Frontiers in Astronomy and Astrophysics, Beijing Normal University, Beijing 102206, China}
\affiliation{School of Physics and Microelectronics, Zhengzhou University, Zhengzhou 450001, China}
\email[]{}

\author[]{Xi Kang}
\affiliation{Institute for Astronomy, the School of Physics, Zhejiang University, Hangzhou 310027, China}
\affiliation{Purple Mountain Observatory, 10 Yuan Hua Road, Nanjing 210034, People’s Republic of China}
\email[]{}

\author[]{Weipeng Lin}
\affiliation{CSST Science Centre for the Guangdong-Hongkong-Macau Greater Bay Area, School of Physics and Astronomy, Sun Yat-sen University, No. 2, DaXue Road, Zhuhai 519082, PR China}
\email[]{}

\author[]{Jie Wang}
\affiliation{Key Laboratory for Computational Astrophysics, National Astronomical Observatories, Chinese Academy of Sciences, Beijing 100101, China}
\affiliation{School of Astronomy and Space Science, University of Chinese Academy of Sciences, Beijing 100049, China}
\email[]{}

% \collaboration{all}{The Terra Mater collaboration}

\begin{abstract}

We introduce a novel method for reconstructing the cosmic mass, tidal, and velocity (MTV) fields over the redshift range $0<z<0.6$ using the phase information of galaxy groups. This approach replaces the explicit theoretical bias correction typically needed to relate galaxy groups to the underlying dark matter density field with a simulation-calibrated statistical mapping, reducing a major source of systematic uncertainty and making the method directly applicable to spectroscopic redshift surveys such as the DESI Bright Galaxy Survey (BGS). We evaluate the performance of our MTV reconstruction pipeline with mock redshift surveys that include a comprehensive set of observational selection effects. The galaxy groups used as tracers are identified with an extended halo-based group finder applied to the DESI mock galaxy catalogue with an apparent magnitude limit of $m_z < 19.65$, yielding a galaxy number comparable to that of the DESI BGS faint sample ($m_r < 20.175$). Our tests show that the reconstructed velocities are accurate and unbiased, with a residual dispersion of $\sim 120\ \mathrm{km\,s^{-1}}$ across the redshift bins. The recovered velocity field allows us to shift galaxy groups to their real-space positions, thereby correcting for the Kaiser effect. By iteratively applying this Kaiser correction to the galaxy groups, we further reconstruct the tidal field and the mass-density distribution. The reconstruction is stable with respect to the grid resolution. Overall, our results demonstrate that this group-based phase-space reconstruction provides a robust pathway to recovering the dark matter MTV fields, with strong prospects for application to DESI BGS data. 
% As MTV reconstruction is a key component of the broader ELUCID-DESI (Exploring the Local Universe with reConstructed Initial Density field using DESI) project, this work is the second paper in the series.

\end{abstract}



\section{Introduction}

As key quantities that describe the large-scale structure of the Universe, the cosmic Mass, Tidal, and Velocity (MTV) fields offer a multiscale characterization of the dark matter distribution, enabling us to probe gravitational dynamics and the evolution of structure \citep{Strauss1995}. Moreover, reliably reconstructing the MTV fields is essential for a broad range of applications, such as modeling and interpreting redshift-space distortions \citep[RSD, e.g.,][]{Zhang2013,Dam2021}, testing gravity on cosmological scales \citep[e.g.,][]{Vittorio1986, Colombi2007, Cai2025}, constraining cosmological parameters like $f\sigma_8$ \citep[e.g.,][]{Pike2005,Shi2018, Lyall2024}, and exploring the relationship between galaxies and their large-scale environments \citep[e.g.,][]{Hoffman2012,Pomarede2017}. Such information may also help connect the large-scale matter and velocity fields to the accretion and angular-momentum history of galaxies, which can play an important role in their growth and quenching \citep[e.g.,][]{Peng2020}.
 
Over the past two decades, numerous techniques have been proposed to reconstruct MTV fields and have been applied to galaxy redshift surveys \citep[e.g.,][]{Fisher1995,Zaroubi1995,Erdogdu2004,Wang2012,Springob2014,Lavaux2016,Yu2019}. An early method based on Wiener filtering was used to recover the cosmic velocity field in the linear regime and was implemented on the 2dF and 2MASS galaxy redshift surveys \citep{Erdogdu2004, Erdogdu2006}. The underlying matter density field can also be reconstructed by weak lensing measurements \citep[e.g.,][]{Amara2012,Jeffrey2021}. In more recent years, machine-learning approaches have been increasingly adopted for MTV field reconstruction \citep[e.g.,][]{Wu2021,Ganeshaiah2023,Wu2023,Chen2023,Wang2024,Wangyy2024,Tanimura2024,deAndres2024,Du2025,Maragliano2025}. As one example, \citet{Shi2025} employed a U-Net convolutional neural network to reconstruct MTV fields from mock realizations of the DESI Bright Galaxy Survey across the redshift interval $0.1 < z < 0.4$.

As research on the galaxy–halo connection has advanced, the tracers used to reconstruct MTV fields have expanded beyond galaxies alone. \citet{Wang2009} proposed a halo-domain technique to reconstruct the mass density field using a catalog of dark matter halos. Furthermore, \citet{Wang2012} demonstrated that dark matter halos or observationally identified galaxy groups can serve as effective tracers for reconstructing velocity and tidal fields, provided a linear bias between halos and the underlying mass density field is assumed. This halo-based reconstruction scheme has been successfully implemented with SDSS data \citep[e.g.,][]{Wang2012,Shi2016,Shi2018}, relying on the galaxy group catalog of \citet{Yang2005,Yang2007}. However, these works primarily target the MTV fields in the nearby Universe ($z < 0.2$), using relatively bright galaxy samples.  

More recently, the Dark Energy Spectroscopic Instrument (DESI), as a fourth-generation galaxy redshift survey, has delivered an unprecedented number of accurate redshift measurements over a much larger sky area and with high completeness out to $z \sim 1$ \citep{Myers_2023, Hahn_2023}. The DESI Bright Galaxy Survey (BGS) has completed its 5-year observing program, mapping more than 15000 $\rm deg^2$ of sky to a significantly fainter limit of $m_r < 20.175$. As a result, the DESI BGS, with its extended redshift coverage and greatly increased survey volume, poses new challenges for MTV field reconstruction, including issues related to structure evolution and redshift-dependent bias.

In this study, we present a new approach to reconstructing the cosmic MTV fields using the phase information of galaxy groups. In contrast to the traditional halo-based reconstruction techniques, our method links galaxy groups to the underlying dark matter distribution through the matter density field in Fourier space, as obtained from simulations, instead of relying on a halo bias model calibrated by the theoretical halo mass function. We use galaxy groups as the main tracers of the underlying matter distribution. From the reconstructed velocity field, we correct for the Kaiser effect and then employ the corrected galaxy group catalogues to recover both the mass density and tidal fields.
% Furthermore, we incorporate a satellite galaxy sample into the mass density reconstruction by adopting a theoretical model to infer subhalo masses. 
We validate our methodology using the DESI mock catalogue, which includes observational selection effects \citep{Gu2024} and is generated from the Jiutian simulation \citep{Han2025}. The mock galaxy sample, selected with an apparent magnitude limit of $m_z < 19.65$, is comparable to the DESI BGS faint sample with $m_r < 20.175$, enabling application of our method to real DESI observations in future work.

%With the reconstructed velocity field, the Kaiser effect on galaxy groups can be effectively corrected, allowing us to reconstruct the mass density field precisely. 
%Among the reconstructed MTV, the mass density field can be employed to recover the initial density field of our Universe and to investigate the growth of structure in the real Universe \citep{Wang2016}. We have carried out the ELUCID (Exploring the Local Universe with reConstructed Initial Density field) simulations \citep[e.g.][]{Wang2014, Tweed2017, Wang2016}, which are constructed from the group catalogs \citep{Yang2007} of the SDSS main galaxy sample and the associated value-added galaxy catalogs \citep{Blanton2005}. This framework enables a one-to-one correspondence between observed galaxies and dark matter subhalos \citep[e.g.][]{Yang2018, Wang2018, Zhang2021, Zhang2022, Zhang2025}. We now perform ELUCID-DESI based on the DESI BGS sample, which provides new opportunities to probe the galaxy–halo connection and to advance our understanding of galaxy evolution. In the first ELUCID-DESI paper \citep{Hong2026}, we implemented an MPI-parallel Bayesian framework with distributed memory to reconstruct initial conditions, thereby overcoming potential scalability limitations of the reconstruction pipeline within a DESI survey volume. As the second paper in the ELUCID-DESI series, the present work concentrates on accurate MTV reconstruction, using a specific mass density field as the input to the MPI initial-condition solver.

Among the reconstructed MTV fields, the mass density field is especially important, as it allows one to infer the initial density field of the Universe and to study the subsequent growth of cosmic structures \citep{Wang2016}. This underpins the ELUCID (Exploring the Local Universe with reConstructed Initial Density field) simulation framework, which has been successfully implemented using SDSS data \citep[e.g.][]{Wang2014, Tweed2017, Wang2016}. By combining group catalogs from the SDSS main galaxy sample \citep{Yang2007} with the corresponding value-added galaxy catalogs \citep{Blanton2005}, ELUCID establishes a one-to-one mapping between observed galaxies and dark matter subhalos, thereby offering a powerful approach to investigate the galaxy–halo connection \citep[e.g.][]{Yang2018, Wang2018, Zhang2021, Zhang2022, Zhang2025}. Extending this framework to the DESI BGS sample opens up new possibilities for exploring galaxy evolution across a much larger cosmological volume and to higher redshifts. In the first ELUCID-DESI paper \citep{Hong2026}, we introduced an MPI-parallel Bayesian framework with distributed memory to reconstruct the initial conditions, thereby overcoming the scalability limitations posed by the DESI survey volume. As the second installment in the ELUCID-DESI series, the present study concentrates on the precise reconstruction of the MTV fields, yielding a reliable mass density field that serves as the crucial input for the initial-condition solver. This establishes the foundation for forthcoming ELUCID-DESI simulations that will facilitate detailed investigations of structure formation, galaxy evolution, and cosmological constraints using the DESI BGS data.

This paper is structured as follows. In Section~\ref{sec:data}, we introduce the mock galaxy sample and the galaxy group catalog derived from the Jiutian simulation. Section~\ref{sec:method} details our methodology for reconstructing the MTV fields, and in Section~\ref{sec:results} we assess the performance of the MTV field reconstruction. Finally, Section~\ref{sec:summary} provides a summary of our key results and conclusions. Throughout, we assume a $\Lambda$CDM cosmology with parameters consistent with the \emph{Planck} 2018 measurements \citep{Planck2020}: $\Omega_\mathrm{m} = 0.315$, $\Omega_{\Lambda} = 0.685$, $n_{\rm s} = 0.965$, $h = H_0/(100\ \rm km\ s^{-1}\ Mpc^{-1}) = 0.674$, and $\sigma_8 = 0.811$.

\section{Data} \label{sec:data}

In this section, we describe the mock galaxy samples and the simulation data employed in this study. We also detail the mock galaxy group catalogue we have constructed, which serves as the primary tracer for reconstructing the MTV fields.

\subsection{Mock galaxy sample}

The mock galaxy light-cone catalogue is constructed based on the Jiutian simulation\footnote{\url{https://jiutian.sjtu.edu.cn/}} \citep{Han2025}, a suite of $N$-body simulations designed to support the planning and scientific analysis of the China Space Station survey Telescope (\emph{CSST}) extra-galactic surveys. The primary run contains three dark matter-only simulations using $6144^3$ particles in boxes of size 0.3, 1, and 2 $h^{-1}$Gpc per side (hereafter, Jiutian-300, Jiutian-1G and Jiutian-2G), adopting the \emph{Planck}-2018 cosmology \citep{Planck2018} with $\Omega_\mathrm{m} = 0.3111$, $\Omega_{\Lambda} = 0.6889$, $n_{\rm s} = 0.9665$, $h = H_0/(100\ \rm km\ s^{-1}\ Mpc^{-1)}=0.6766$ and $\sigma_8=0.8102$. The corresponding particle masses are $1.005 \times 10^7\ h^{-1}\mathrm{M_{\odot}}$, $3.723 \times 10^8\ h^{-1}\mathrm{M_{\odot}}$ and $2.978 \times 10^9\ h^{-1}\mathrm{M_{\odot}}$ for the Jiutian-300, Jiutian-1G and Jiutian-2G, respectively. Jiutian-1G was run using L\textsc{Gadget-3} \citep{Angulo2012}, while Jiutian-300 and Jiutian-2G were run with the newly released \textsc{Gadget-4} code \citep{Springel2021}. Each simulation outputs 128 snapshots from an initial redshift $z = 127$ to $z = 0$. 

Dark matter halos are identified with a friends-of-friends (FoF) algorithm \citep{Davis1985} using a linking length of 0.2 times mean interparticle separation. Subhalos and merger trees are constructed using two independent pipelines. The first employs \textsc{Subfind} \citep{Springel2001}, a configuration-space subhalo finder that identifies self-bound, locally overdense substructures within FoF halos. The second uses \textsc{HBT+} \citep{Han2012,Han2018}, a time-domain subhalo finder that tracks the evolution of each halo across snapshots to build hierarchical merger trees.

A light-cone halo/subhalo catalogue is constructed from Jiutian-1G according to the evoluation of FoF halos and \textsc{HBT+} subhalos. An observer is placed at a reference location in the origin box, and the simulation boxes are periodically replicated to fill the light-cone volume.  Subhalos are constructed in sequence from the snapshots, following the comoving distances and redshifts at which they are observed.  Spectroscopic redshifts are derived from the comoving distances of galaxies, while also incorporating their line-of-sight peculiar velocities. In total we make use of 46 snapshots from $z = 0$ to $z = 1.03$  to build the light cone. In addition, we include a typical redshift uncertainty of $35\ \rm km\,s^{-1}$, representative of a standard spectroscopic redshift survey such as DESI. Then, galaxy luminosities are assigned to subhalo populations using the subhalo abundance matching (SHAM) method with DESI $z$-band luminosity functions at different redshifts. The DESI Legacy Survey (LS) geometry and magnitude limit cuts are applied. The north Galactic cap (NGC) and south Galactic cap (SGC) regions of the DESI LS DR9 footprint cover a total sky area of approximately 18,350 $\rm deg^2$ \citep[see Figure 4 in][]{Gu2024}. 

%To incorporate RSD in the mock galaxy catalog, we adopt the plane-parallel (distant observer) approximation, in which the line-of-sight (LOS) direction is assumed to be aligned with the z-axis of the simulation box. Under this assumption, the real-space positions of galaxies are shifted along the z-direction according to their peculiar velocities. Specifically, the redshift-space coordinate is given by
%\begin{equation}
%    s_{\rm z} = \mathrm{z} + \frac{v_{\rm z}}{aH(z)},
%\end{equation}
%where z is the real-space position, $v_{\rm z}$ is the peculiar velocity component along the z-axis, $a$ is the scale factor, and $H$ is the Hubble parameter at the corresponding redshift.

As a method test, we only consider NGC region in this work that covers a region of 9621.98 $\rm deg^2$. The total number of mock galaxy sample is 16,110,660 selected with $m_z < 19.65$ and $z < 0.6$. Specially, our mock galaxy sample contains a number of 12,450,419 and 3,660,241 galaxies at $0 < z \leq 0.4$ and $0.4 < z < 0.6$, with a number density of $\sim 1.0 \times10^{-2}$ and $1.6 \times 10^{-3}\ h^{3}\rm Mpc^{-3}$, respectively. This mock galaxy sample is comparable to the DESI BGS faint sample with $m_r < 20.175$.

\subsection{Mock galaxy group catalog}

\begin{figure}
    \centering
    \includegraphics[width=0.45\textwidth]{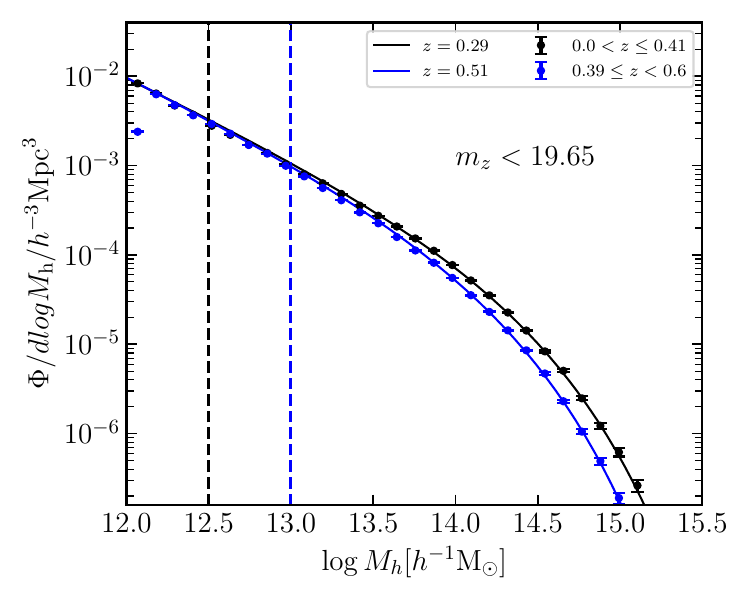}
    \caption{The halo mass function of mock galaxy groups catalogue, separated in two redshift bins. The error bars represent Poisson uncertainties. The black and blue vertical dashed lines mark the position of $10^{12.5}$ and $10^{13} \Msunh$, representing the mass limit of galaxy groups adopted for the velocity field reconstruction at $0.0 < z \leq 0.41$ and $0.39\leq z < 0.6$. The solid lines show the theoretical halo mass function predicted by \citet{Sheth2001}.  %\adr{XXX adopt lower mass cut for low redshift bin, e.g., 11.5?  XXX } \qy{Though the HMF seems complete at $M_h > 10^{11.5} \Msunh$ for the groups at $0.0 \leq z \leq 0.4$, however, the HMF at $0.2 \leq z \leq 0.4$ drops at $M_h \sim 10^{11.6} \Msunh$. This has also been checked in the velocity reconstruction. The components of velocity are biased when adopting groups with $M_h > 10^{11.5} \Msunh$. This situation also suits for the groups at $0.4 \leq z \leq 0.6$.}
    }
    \label{fig:hmf}
\end{figure}

% \begin{figure*}
%     \centering
%     \includegraphics[width=0.98\textwidth]{figs/grpmap.pdf}
%     \caption{Examples slices of the spatial distribution of galaxy groups identified with \zspec\ (left panel) and \zphoto\ (right panel). The slice is cut with $\rm \Delta Dec = 0.1$ at $\rm RA = 40\ deg$. The mass criteria of groups at different redshift ranges causes the evolution in group number density with redshift.}
%     \label{fig:grpmap}
% \end{figure*}

The mock galaxy group catalog is constructed from the mock galaxy sample as described in the previous section. Galaxy groups are identified using a halo-based group finder originally developed by \citet{Yang2005,Yang2007}. In this work, we adopt the extended version of the halo-based group finder \citep{Yang2021}, which is applicable to both spectroscopic and photometric redshift samples. The halo properties including halo radius, velocity dispersion, as well as galaxy-halo connections (e.g.,  stellar to halo mass ratios) together provide the foundation of this group finder. The extended halo-based group finder has been successfully applied to the observations, including DESI \citep{Yang2021, WangYR2024}, CLAUDS and HSC \citep{Li2022}, and COSMOS2020 galaxy catalog \citep{Li2025}. We briefly summarize the main steps of this method; a detailed description can be found in \citet{Yang2021}.

The group-finding procedure begins by treating each galaxy as a tentative group. Halo masses are assigned on the basis of mass-to-light ratios, calibrated through interpolation. The mass-to-light ratios of the groups in each redshift bin are determined with cumulative halo mass functions \citep{Sheth2001} and the group luminosity or stellar mass functions using the abundance matching method \citep{Yang2007}.
Given the halo mass, each group is assigned a halo radius and velocity dispersion along the line-of-sight. The group membership updates begin from the most massive one by taking the luminosity-weighted group center as the halo center and assuming that the distribution of member galaxies in coordinate and velocity spaces follows that of the dark matter particles or subhalos. Galaxies are assigned to a candidate group with a judgment of the number density contrast of galaxies in the redshift space around the group center to that at the edge of a halo. 
After assigning all the galaxies into groups, we update the group centers and luminosities, recalculate the halo information, and find member galaxies again. The iteration stops until there are no more changes for the group memberships. Finally, we start from the beginning to make another iteration, aiming at the convergence of mass-to-light ratios.
After applying the extended halo-based group finder, we obtain 12,042,518 galaxy groups in the mock NGC region. In the running of the group finder, a $K$-correction with a formula of $K_{\mathrm{z}}^{0.5}(z) = -0.35-0.56z+0.81z^2$ is also applied to apparent magnitude to obtain a consistent luminosity.

In the reconstruction, we divide the galaxy group samples into two redshift bins. 
For the velocity field reconstruction, we consider the redshift ranges $0.0 < z \leq 0.41$ and $0.39 \leq z < 0.6$, with the lowest-redshift bin corresponding to the DESI Bright Galaxy Survey range. Since our primary focus is on galaxy groups in the two redshift intervals $0.0 < z < 0.4$ and $0.4 < z < 0.6$, using slightly broader redshift ranges for the velocity field reconstruction helps reduce boundary effects from the survey, leading to more reliable field estimates for groups located near the edges of these redshift bins. 
% The interlaced redshift bins are adopted to avoid the effect of the survey boundary on field reconstruction. 
In Figure~\ref{fig:hmf}, we show the halo mass function (HMF) of the mock galaxy groups in the two redshift bins. Taking into account the survey magnitude limit and the completeness of the group, we adopt a cutoff halo mass of $10^{12.5} \Msunh$ and $10^{13}\Msunh$ for two redshift bins in the field reconstruction, as indicated by dashed vertical lines.
% \adg{Although the HMF seems complete at a smaller halo mass, for example $M_h > 10^{12.5} \Msunh$ for the groups at $0.39 \leq z < 0.6$, the HMF in a higher redshift range (e.g., $0.5 < z < 0.6$) becomes incomplete at $M_h \sim 10^{12.5} \Msunh$.} 

Here we select two sets of galaxy and group samples for our MTV field reconstruction. First, for the velocity field reconstruction, we apply a conservative halo mass threshold to ensure a complete group sample across the full redshift range. The motivation for these requirements will be discussed later in section \ref{sec:vel}.
The information of the galaxy groups and the reconstruction box for the velocity field in the two redshift bins is listed in Table~\ref{tab:box}. For the other set, used for reconstructing the mass density and tidal field, we employ all galaxies and groups with non-overlapping redshift ranges, with the relevant information summarized in Table.~\ref{tab:box_dens}.  Since the mass density field will serve as input for future ELUCID-DESI simulations, we adopted a rotated survey region embedded within a cubic box to reduce the box size used in our reconstruction.

% Figure~\ref{fig:grpmap} shows a slice of the spatial distribution of galaxy groups in different redshift ranges. The \zspec\ groups distribution clearly traces cosmic web structures, while becoming increasingly sparse at higher redshift due to the mass limit of the sample.

\begin{table*}
    \centering
    \caption{Information of the fiducial reconstruction box for velocity field reconstruction. The first column lists the different redshift bins. We also list the side size of box ($L_{\rm box}$) in unit of $h^{-1}\rm Mpc$, number of total grids used in reconstruction ($N_g$), group mass limitation ($M_{\rm th}$) in unit of $h^{-1}\mathrm{M_{\odot}}$, total number of groups used in reconstruction ($N_{\rm grp}$), number density of groups ($n_{\rm grp}$) in unit of $h^{3}\rm Mpc^{-3}$, and volume-weighted redshift ($z_{\rm eff}$).
    }
    \label{tab:box}
    \hspace{-1.2cm}
    \begin{tabular}{c c c c c c c}
    \hline\hline
    redshift & $L_{\rm box}$  & $N_g$ & $M_{\rm th}$ & $N_{\rm grp}$ & $n_{\rm grp}$ & $z_{\rm eff}$ \\ 
    % & $h^{-1}\mathrm{Mpc}$ & & $h^{-1}\mathrm{Mpc}$ & $h^{-1}\mathrm{M_{\odot}}$ & $10^9h^{-3}\mathrm{Mpc}^3$\\
     \hline
     $0.0 < z \leq 0.41$ & 2064 & $512^3$ &  $10^{12.5}$ & 1,680,773 & $1.4\times10^{-3}$ & 0.29\\
     $0.39 \leq z < 0.6$ & 2822 & $512^3$ &  $10^{13.0}$ & 763,470 & $3.3\times10^{-4}$ & 0.51\\
    \hline
\end{tabular}
\end{table*}

\begin{table*}
    \centering
    \caption{Similar to Table~\ref{tab:box} but for the dark matter mass density and tidal field reconstruction. The redshift ranges are restricted to $0.0 < z < 0.4$ and $0.4 < z < 0.6$ with a rotated survey volume. We do not adopt a group mass limitation in the reconstruction where the minimal group mass are $\sim 10^{10.55}$ and $10^{12.12} \Msunh$.
    }
    \label{tab:box_dens}
    \hspace{-1.2cm}
    \begin{tabular}{c c c c c c c}
    \hline\hline
    redshift & $L_{\rm box}$  & $N_g$ & $M_{\rm th}$ & $N_{\rm grp}$ & $n_{\rm grp}$ & $z_{\rm eff}$ \\ 
     \hline
     $0.0 < z < 0.4$ & 1680 & $512^3$ & - & 9,030,603 & $7.3\times10^{-3}$ & 0.29\\
     $0.4 < z < 0.6$ & 2310 & $512^3$ & - & 3,014,122 & $1.3\times10^{-3}$ & 0.51\\
    \hline
\end{tabular}
\end{table*}

\begin{figure*}
    \centering
    \includegraphics[width=1.0\textwidth]{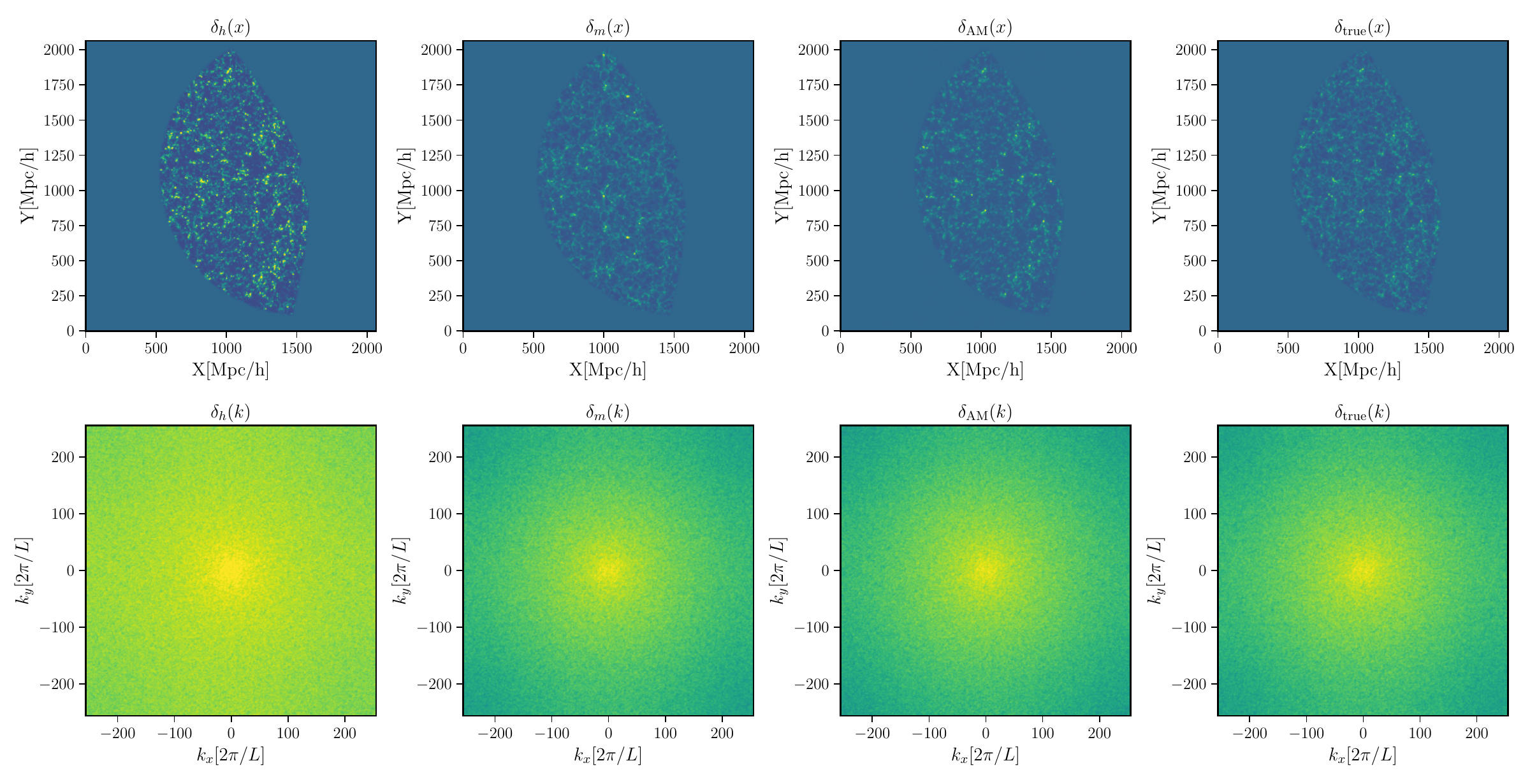}
    \caption{The mass density field in survey box and in corresponding phase-space at $0.0 < z < 0.41$. The top panels show a slice of density field along $Z-$axis with thickness of $32.25 \mpch$, while the bottom panels present a slice of the amplitude of phase-space at $k_x$-$k_y$ plane. The first, second, third, and fourth columns represent the galaxy groups mass density field, the independent matter density field used for phase-space abundance matching, the group mass density field after phase-space abundance matching, and true mass density field, respectively.}
    \label{fig:phase_space}
\end{figure*}

\section{Method} \label{sec:method}

In this section, we introduce the method for reconstructing the cosmic MTV fields using galaxy group catalog. We begin by outlining how we use the phase information of galaxy groups to perform our MTV reconstruction. Then, we describe the procedure adopted for reconstructing the velocity field.  
Subsequently, we detail the method used to recover the mass density field from galaxy groups, where the Kaiser effect is corrected by means of the reconstructed velocity field. Finally, we present the method for the reconstruction of tidal field relying on the reconstructed mass density field.  

\subsection{phase-space abundance matching} \label{sec:phase-space}

The key step in reconstructing the MTV fields using galaxy groups is to link the spatial distribution of these groups to the underlying dark matter structure. In previous studies, this connection was described by the halo model through a constant halo bias factor  \citep{Wang2012}. However, halo bias models rely on different assumptions that may not accurately represent a specific halo sample, potentially introducing systematic biases. Moreover, halo bias is typically measured from the correlation function or power spectrum on large scales and therefore does not fully capture the small-scale information. To overcome these limitations, we develop a new approach that links the galaxy group distribution to the underlying dark matter distribution using the phase information of the galaxy groups, without relying on an explicit halo bias model. The central assumption we make is that the halo matter density field inferred from the group distribution, $\delta_h(\bm{x})$, encodes the same phase information as the actual dark matter density field as shown in the Appendix~\ref{sec:phase}, $\delta_m(\bm{x})$, differing only by overall amplitude factors.

We transform $\delta_h(\bm{x})$ into $\delta_m(\bm{x})$ in Fourier space following the steps outlined below.
First, we reconstruct the halo mass density field, $\delta_h(\bm{x})$, based on the spatial distribution of galaxy groups. Here $\delta_h(\bm{x})$ is defined as,
\begin{equation}
    \delta_{\rm h}(\bm{x}) = \frac{\rho_{{\rm h}}(\bm{x})}{\bar{\rho}_{\rm h}} - 1 \,,
\end{equation}
where $\rho_{{\rm h}}(\bm{x})$ is the halo mass density calculated using the galaxy groups and $\bar{\rho}_{\rm h}$ is the mean mass density of groups within the survey volume that is defined as the region occupied by groups in survey box. Note that, as will be detailed in the following subsections, we may adopt different methods to compute $\bar{\rho}_{\rm h}$ for the reconstruction of the velocity field and of the mass density field, respectively.
The $\delta_h(\bm{x})$ is then converted to Fourier form using Fourier transform with an expression of $\delta_h(\bm{k})$. 

At the same time, we constructed a dark matter mass density field, $\delta_m(\bm{x})$, based on a cosmological simulation (Jiutian-1G in this work) and the field in Fourier space, $\delta_m(\bm{k})$. The snapshot in simulations with redshift corresponding to $z_{\rm eff}$ is used to construct DM density field. The $\delta_m(\bm{x})$ has the same survey box information as the box to contain galaxy groups, including the box size, number of grids, smoothing scale, and the survey boundary. The only difference is the spatial distribution between galaxy groups and dark matter in the survey region. This means that both density fields have the same survey effects, which allows us to focus on only the intrinsic halo assembly differences. 

Next, we keep the phase of $\delta_h(\bm{k})$ fixed and replace its amplitude $|\delta_h(\bm{k})|$ with the amplitude $|\delta_m(\bm{k})|$ using an abundance-matching procedure. Concretely, we first sort $|\delta_h(\bm{k})|$ into bins according to the wavenumber $k$, and do the same for $|\delta_m(\bm{k})|$. Within each $k$-bin, we rank the values of $|\delta_h(\bm{k})|$ and $|\delta_m(\bm{k})|$, then replace each $|\delta_h(\bm{k})|$ by the $|\delta_m(\bm{k})|$ that has the same rank order. In this way, we complete the mapping from the galaxy group distribution to the underlying dark matter distribution in Fourier space and obtain $\delta_{\rm AM}(\bm{k})$, which has the amplitude of $\delta_m(\bm{k})$ but the phase of $\delta_h(\bm{k})$. Finally, we apply the Fourier transform again and obtain the underlying matter distribution in real-space.

As an example, the upper panels of Figure~\ref{fig:phase_space} show, from left to right, the galaxy group density field, an independent matter density field, the reconstructed matter density which is obtained by preserving the phase information while rescaling the amplitude according to the independent matter density field, and the true matter density field. Here, the independent mass density field is constructed from the same Jiutian-1G simulation but at a different location. We have also verified that adopting a different density field, e.g., one derived from the Jiutian-2G simulation, as the independent mass density field does not affect results. The locations of galaxy groups have been shifted to their true real-space positions by properly accounting for large-scale redshift-space distortion effects, based on the velocity reconstruction that will be described in detail in Section \ref{sec:vel}.
%Within this framework, It is consistent with the usage of phase-space information that the invariable phase information keep the matter geometry distribution unchanged, while the amplitude information also controls the amplitude of mass density fields.   
In the bottom panels of Figure~\ref{fig:phase_space}, we show the slice of the amplitude of phase-space in different density fields that are used in the velocity field reconstruction. The amplitude of the phase-space after amplitude abundance matching (middle panels) is obviously decreased compared with the intrinsic amplitude of phase-space for the density field of galaxy groups (left panels).

In contrast to the density field reconstruction approach of \citet{Wang2012}, which presumes a constant proportionality between the halo and dark matter density fields,
$b_{\rm hm} = \delta_{\rm h}(\bm{x})/\delta(\bm{x})$, understood as the mass-weighted mean halo bias,
\begin{equation} \label{eq:theory_bhm}
    b_{\rm hm} = \frac{\int_{M_{\rm th}}^{\infty} M\, b_{\rm h}(M)\, n(M)\, dM}{\int_{M_{\rm th}}^{\infty} M\, n(M)\, dM}\,,
\end{equation}
our phase-space reconstruction exhibits differences in both its overall amplitude and its scatter. In this expression, $n(M)$ and $b_{\rm h}(M)$ denote the halo mass function and the halo bias function, respectively. While $b_{\rm h}$ can be readily computed within the halo model framework, we observe that various theoretical models for the halo bias $b_{\rm h}$ differ substantially, especially at high redshift and for high halo mass thresholds $M_{\rm th}$ as shown in Appendix~\ref{sec:bhm}.

In Figure~\ref{fig:bhm}, we show the median of $b_{hm}(k)\equiv \delta_{h}(k)/\delta_{m}(k)$ obtained from our phase-space abundance matching method. On large scales, $b_{hm}(k)$ remains approximately constant, but it departs from a linear bias and grows markedly toward smaller scales ($k \gtrsim 0.1\ h\rm Mpc^{-1}$). This rise occurs because the halo density field is traced only by relatively massive halos, whose one-halo term is far more pronounced than that of the dark matter particles. As the phase-space abundance matching procedure effectively imposes a deterministic, mode-by-mode correspondence between the halo and matter density fields, the resulting reconstruction naturally preserves this scale-dependent nonlinear halo bias.
When we compare these results with various halo model predictions in the literature, we find that only a subset of them accurately reproduce the large-scale behavior. In practice, the inappropriate single value of $b_{hm}$ used in cosmic field reconstruction can significantly affect the results of reconstruction, for example, biasing the relation between the reconstructed and true velocity.

\begin{figure}
    \centering
    \includegraphics[width=0.45\textwidth]{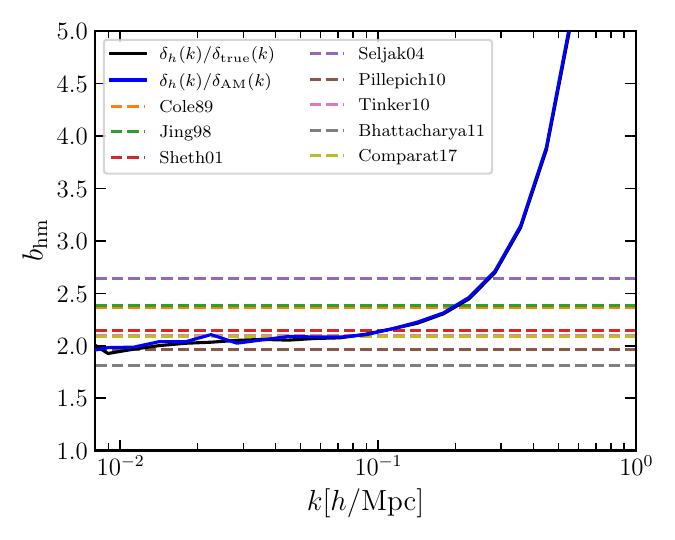}
    \caption{The distribution of $b_{\rm hm}$ as a function of phase scale $k$. The black and blue lines represent the mean value of the ratio between the amplitude of galaxy group density field in Fourier space and the amplitude of true density field and the density field after phase-space abundance matching. The lines with different colors represent the bias parameters derived from different theoretical models, including \citet{Cole1989}, \citet{Jing1998}, \citet{Sheth2001}, \citet{Seljak2004}, \citet{Pillepich2010}, \citet{Tinker2010}, \citet{Bhattacharya2011}, and \citet{Comparat2017}.}
    \label{fig:bhm}
\end{figure}

\subsection{Velocity field} \label{sec:vel}

Using the phase-space reconstruction algorithm described above, we begin by presenting our approach to reconstructing the velocity field. The derivation of the velocity field from galaxy groups is based on the continuity and Poisson equations, along with the fundamental relation that links dark matter to dark matter halos \citep[e.g.,][]{Mo2010}. This galaxy group-based method was first explored by \citet{Wang2012}, who successfully applied it to galaxy group samples selected from the SDSS with redshifts in the range $0.01 \leq z \leq 0.12$. Our method for building the dark matter density field is based on the general framework of \citet{Wang2012}, with a crucial modification: we rely on the phase information of galaxy groups, while the overall amplitude and scatter are taken directly from a specified dark matter simulation. In the following subsections, we elaborate on the theoretical foundations and detailed reconstruction procedures adopted in this work.

\subsubsection{Basis of theory}

In the linear regime, combined with the Poisson equation and the peculiar velocity derived from the peculiar gravitational potential, the velocity field in Fourier space can be written as:
\begin{equation} \label{eq:vfield}
    \bm{v}(\bm{k}) = Haf(\Omega_m)\frac{i\bm{k}}{k^2}\bm{\delta}(\bm{k}),
\end{equation}
where $\bm{v}(\bm{k})$ and $\delta(\bm{k})$ are the Fourier transforms of velocity and matter density contrast field, respectively. The linear growth rate is defined as $f(\Omega_m) = \mathrm{d}\ln D/ \mathrm{d} \ln a$, and is well approximated by
\begin{equation}
    f(\Omega_m) \simeq \Omega_m^{0.6} + \frac{1}{70}\Omega_{\Lambda}\left(1+\frac{\Omega_m}{2}\right),
\end{equation}
\citep[e.g.,][]{Lahav1991}. We note that the quantities $H$, $a$, and $f(\Omega_m)$ are all redshift-dependent and are evaluated at the volume-weighted redshift of each reconstruction volume, $z_{\rm eff}$. The linear velocity field $\bm{v}(\bm{x})$ is obtained by taking the Fourier transform of $\bm{v}(\bm{k})$.

\begin{figure*}
    \centering
    \includegraphics[width=0.90\linewidth]{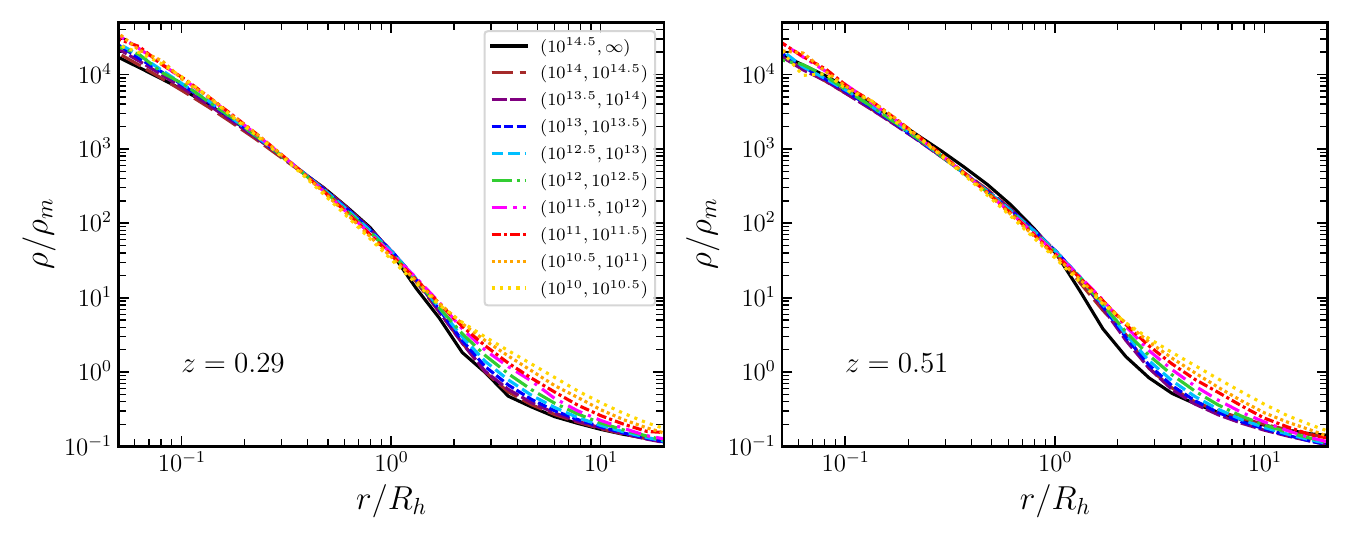}
    \caption{Density profile from halo-domain particles in different halo mass bins at $z = 0.29$ and $z = 0.51$. The density and radius are rescaled to the mean mass density of the universe and virial radius, respectively. The halo mass range shown in the legend is in unit of $\Msunh$. The halo mass range of halo population in determining the domain region is $M_h > 10^{10}\Msunh$.  
    % \adr{XXX make two panels, at redshifts 0.29 and 0.51, respectively. add lower mass bin results as well. XXX }
    }
    \label{fig:dmdens}
\end{figure*}

\subsubsection{Reconstruction Procedure}

In both the observations and our realistic mock galaxy and group sample, where the light-cone effect is included, dark matter halos and groups at different redshifts correspond to structures that undergo evolution. However, the velocity reconstruction using Eq. \ref{eq:vfield} requires $\bm{\delta}(\bm{k})$ evaluated at a single, fixed redshift. We begin our reconstruction of the velocity field by selecting galaxy groups within a given redshift range and then rescaling their halo masses to a chosen effective redshift $z_{\rm eff}$, thereby mitigating the impact of structure growth. This rescaling is performed through abundance matching, using halo mass functions derived from the Jiutian simulation at multiple snapshots and redshifts.

In addition to correcting for evolutionary effects, to minimize magnitude-related selection biases in the survey, we restrict our sample to galaxy groups with halo masses $M_{\rm h} > M_{\rm th}$ and place them within a cubic volume (hereafter the survey box). The sky coordinates (RA, Dec, $z$) of galaxy groups are converted into Cartesian coordinates $(X, Y, Z)$, with the midpoint of each axis placed at the centre of the survey box. The survey box is divided into uniform grid cells. The information of selection criteria including survey box, number of groups, etc., for the reconstruction of velocity field is listed in Table.~\ref{tab:box}. Then, the groups are assigned to the grids to construct the halo mass density field, $\rho_{\rm h}$. Following \citet{Wang2012}, the mass of each group is homogeneously distributed within a spherical region of radius $R_{200}/2$ around the group centre. The $R_{200}$ radius is defined as the radius within which the mean enclosed density is 200 times the mean density of the universe at the redshift of the group. The total density contrast fields of galaxy groups and dark matter particles, $\delta_{\rm h}$ and $\delta_{\rm m}$, are calculated following the procedure outlined in Section \ref{sec:phase-space}. The survey volume is defined using the Python package \texttt{shapely}, and all grid cells lying outside this region are assigned a value of zero. To obtain $\delta_{\rm h}(\bm{x})$, we apply Gaussian smoothing with a kernel whose width is set to half the size of a grid cell.
% \begin{equation}
%     \sigma_s = \left( \frac{3M_s}{4\pi \bar{\rho}_h} \right)^{1/3},
% \end{equation}
% where we adopt a smoothing mass of $M_s = 10^{14.75} h^{-1}\rm M_{\odot}$ as same adopted in \citet{Wang2012}.

Next, we use the smoothed dark matter density field in Fourier space to estimate the velocity field in Fourier space, $\bm{v}(\bm{k})$, following Equation~\ref{eq:vfield}. The corresponding real-space velocity field, $\bm{v}(\bm{x})$, is then derived by performing an inverse FFT on $\bm{v}(\bm{k})$. This reconstructed peculiar velocity field is finally employed to correct the linear redshift-space distortion and relocate galaxy groups to their true positions in real-space. Here, the cosmological redshift of the groups is calculated as
\begin{equation} \label{eq:zcorr}
    z_{\rm cos} = \frac{z_{\rm obs} - (v_{\rm los}/c)}{1 + (v_{\rm los}/c)},
\end{equation}
where $v_{\rm los}$ denotes the inferred line-of-sight component of the peculiar velocity obtained from $\bm{v}(\bm{x})$, $z_{\rm obs}$ is the observed group redshift, and $c$ is the speed of light. Since the velocity field is initially reconstructed from the redshift-space distribution of galaxy groups, the method is inherently iterative. We therefore repeat the reconstruction, each time updating the group positions to their cosmological redshifts $z_{\rm cos}$, until a stable solution is reached. We perform four iterations, which we find to be sufficient for convergence. Finally, the peculiar velocity field $\bm{v}(\bm{x})$ is obtained from the galaxy groups with these corrected redshifts $z_{\rm cos}$, using the same reconstruction method described above. In our analysis, we keep both the initial and final velocity fields $\bm{v}(\bm{x})$, derived respectively from the redshift-space and real-space group distributions.

\subsection{Mass density field} \label{sec:method-mass}

With the reconstructed linear velocity field in hand, we proceed to construct the nonlinear density field. To make full use of the constraining power of the observations, we include all groups in our sample that fall within the relevant redshift interval in the reconstruction.  Before carrying out the main steps of the mass density field reconstruction, we use the velocity field obtained in the previous subsection to iteratively estimate $z_{\rm cos}$ for the groups that are not involved in the velocity field reconstruction. Meanwhile, we rescale the mass of galaxy groups to the specific redshift $z_{\rm eff}$ through abundance matching, as the operation adopted in the velocity field reconstruction. The corresponding information on the redshift range, box size, number of groups, and related quantities used in our subsequent density field reconstruction is summarized in Table \ref{tab:box_dens}.

We reconstruct two kinds of mass density fields. The first is the Halo-Domain Density Field (HDDF), obtained from galaxy groups using the halo-domain method introduced by \citep{Wang2009}. The second is the Environment-Enhanced Density Field (EEDF), which extends the HDDF by incorporating the contribution from the surrounding large-scale environment. For both types of reconstructions, we additionally apply phase-space abundance matching to better align the reconstructed mass distribution with the true one. The two reconstruction methods are detailed below.

\subsubsection{Halo-Domain Density Field}

The HDDF is reconstructed by assigning galaxy groups to the survey volume, with weights given by the masses of their host halos, following halo-domain density profiles \citep{Wang2009}. The domain of a given halo is defined such that every dark matter sample particle within it is closer to that halo (under a distance metric specified below) than to any other halo in the population. Following \citet{Wang2009}, we adopt a distance ratio to quantify the domain region, expressed as: 
\begin{equation}
    r_{\rm d} = \frac{r_h}{R_h},
\end{equation}
where $r_h$ is the distance between the halo centre and the sample particle and $R_h$ is the halo virial radius adopted as $R_{200}$ here.
Using this halo-domain approach, one adopts a halo-domain density profile to sample particles that are then used to compute the mass density field. Consequently, we first determine the halo-domain density profiles from the Jiutian-1G and Jiutian-300 simulations. In detail, a population of halos with $M_h > 10^{10} \Msunh$ are selected from both simulations at first. The halo-domain density profile for halos with $M_h > 10^{12} \Msunh$ is derived from the Jiutian-1G simulation, using five cubic subvolumes of side length $100\mpch$ to secure a sufficient number of massive halos. For halos with $10^{10} < M_h \leq 10^{12} \Msunh$, the profile is obtained from the Jiutian-300 simulation, based on a single cubic subvolumes of side length $50\mpch$, which provides an adequate particle sampling for low-mass halos. 
% In the subvolumes of both simulations, only halos with $M_h > 10^{10} \Msunh$ are included in constructing the density profiles. 

Figure~\ref{fig:dmdens} displays the mean halo-domain density profiles for various halo mass intervals, at the volume-weighted effective redshift for the two redshift bins. Within the virial radius, the density profiles follow NFW forms, while beyond $\sim 2 R_{h}$ they decline slightly more with increasing halo mass, reflecting the enhanced infall around more massive halos \citep[e.g.,][]{Fong2021}, in agreement with \citet{Wang2009}. In addition, at lower redshift the inner profiles ($\lesssim 0.1 R_h$) are marginally steeper, a consequence of mass growth which increases the mass but barely affects the inner profile \citep[e.g.,][]{Gao2023}. Therefore, a low-redshift halo profile can be mapped to a lower-mass halo profile at high-redshift, which has a slightly steeper inner slope. Relative to the mass-assignment scheme used in reconstructing the velocity field, the halo-domain method extends the effective halo boundary to $20$ times the virial radius, thereby incorporating the two-halo term.

%In the mass assignment, we also include a satellite galaxy population in order to enhance the description for small scales or substructures. The mass of satellite galaxies is derived from the satellite luminosity using a theoretical fitting formula (Li et al. in prep, more details in Appendix.~\ref{sec:submass}). However, the galaxies suffer from the Finger-of-God (FoG) effect even after correcting the redshift with reconstructed velocity field, while we eliminate the FoG effect based on the position of galaxies perpendicular to the line of sight. We calculate the separations between satellite galaxies and its group center parallel and perpendicular to the line-of-sight, $r_\pi$ and $r_p$. Then, the $r_\pi$ for each satellite is randomly assigned according to the probability density distribution of $r_p$. The new redshift of satellite galaxies is derived from the distance of group center plus the reassigned $r_p$. 

Once the mass density fields have been derived using the halo-domain method, we carry out phase-space abundance matching to obtain the final DM density fields, following the procedures described in Section \ref{sec:phase-space}. Finally, we reconstruct the DM density field by applying a Fourier transform to the phase-space abundance–matched density field.

\subsubsection{Environment-Enhanced Density Field}

As we will demonstrate below, within the HDDF framework the reconstructed density field is trustworthy only in regions populated by halos, whereas filamentary and sheet-like (pancake) structures are inadequately represented. To remedy this, we propose an alternative density-field reconstruction method, EEDF, which is built using both halo-domain and environmental DM particles. In practice, after creating a sample of DM particles linked to galaxy groups, we add an extra population of DM particles to better capture and describe the cosmic web. As noted by \citep{Fang2024}, incorporating information from the cosmic web can enhance the reconstruction of the density field.

Here, we utilize Jiutian-1G simulation as a reference to calculate the background density of large-scale environments. For the redshift bin $0.0 < z < 0.4$, we choose the snapshot close to its $z_{\rm eff}$. The DM particles in the Jiutian-1G snapshot are assigned to the $512^3$ grids using the CIC method smoothed with a Gaussian sigma of 0.5 grid cells. Then we compute the tidal field and the eigenvalues of the tidal tensor based on the resulting DM density field (see details in Section.~\ref{sec:tidal}). The grid cells from DM density field is divided into four types of large-scale environments: \emph{void}, \emph{sheet}, \emph{filament}, and \emph{cluster}, according to the eigenvalues of tidal tensor. For each environment, we estimate the background mass by subtracting the mass contained in halos with $M_h > 10^{10} \Msunh$ from the total DM mass. The background density, $\rho_{\rm bg}$, is then obtained by dividing the background mass by the corresponding environmental volume.

Based on the HDDF, we first assign a large-scale environmental type to each grid cell. Given this environmental classification, we generate background DM particles with a spatially uniform distribution inside each cell, where the particle number density is set by the corresponding background density, $\rho_{\rm bg}$. We then use both halo-domain particles and environmental background DM particles to compute the density field in each grid cell  and apply the phase-space abundance matching technique to derive the EEDF.

\subsubsection{Summary of reconstruction procedure}

Next, we outline the primary steps involved in reconstructing the mass density field as follows:  
\begin{enumerate}
    \item Select a sample of galaxy groups in a given redshift interval.
    \item Correct Kaiser effect of galaxy groups using the reconstructed velocity field.
    \item Rescale the halo mass of galaxy groups to the redshift $z_{\rm eff}$ through abundance matching method.
    %\item Reassign the position of satellite galaxies along line of sight according to the probability density distribution of $r_p$.
    \item Rotate the sky coordinate of groups, i.e. $\rm RA - 58.0$, and embed them in a cubic box.
    \item Select, for each group of halos \(h_i\), the corresponding average halo-domain density profile whose mass falls within the mass bin of that profile, and then use a Monte Carlo method to sample particles \(S_h\) according to this density profile in a region extending to roughly 20 times the virial radius of halo \(h_i\).
    \item Verify whether the distance $r_{\rm d}$ between the particle and the center of group $h_i$ is less than its distance to all other groups. If this condition is met, the particle is kept; if not, it is discarded. 
    \item Repeat steps 5 and 6 for each of the selected group populations. 
    \item Construct temporary DM density field with galaxy groups using phase-space abundance matching technique and classify the large-scale environments.
    \item Sample DM particles in every grid cell based on the local environment type and the corresponding background density of that environment.
    \item Build the mass density field using all sampled particles with the Cloud-In-Cell (CIC) method and smooth with a Gaussian kernel.
    \item Replace the amplitude of the reconstructed mass density field in Fourier space in step 10 with the amplitude of a DM density field.
\end{enumerate}

The reconstruction of HDDF includes all procedures except steps 8 and 9, whereas EEDF includes all steps. In summary, we generate two types of DM particles: halo-domain particles generated in steps 5–7, and large-scale environment background particles generated in steps 8 and 9. Therefore, EEDF contains both types of particles, while HDDF only includes halo-domain particles.  
In our implementation, we adopt a halo mass threshold of $M_{\rm th} = 10^{10}\Msunh$, which includes all galaxy groups in the group catalog. 
In addition, we rotate the sky distribution of galaxy groups to embed the survey volume within a smaller simulation box, thereby significantly reducing the memory requirements for the future ELUCID-DESI reconstruction.

\begin{figure*}
    \centering
    \includegraphics[width=0.75\linewidth]{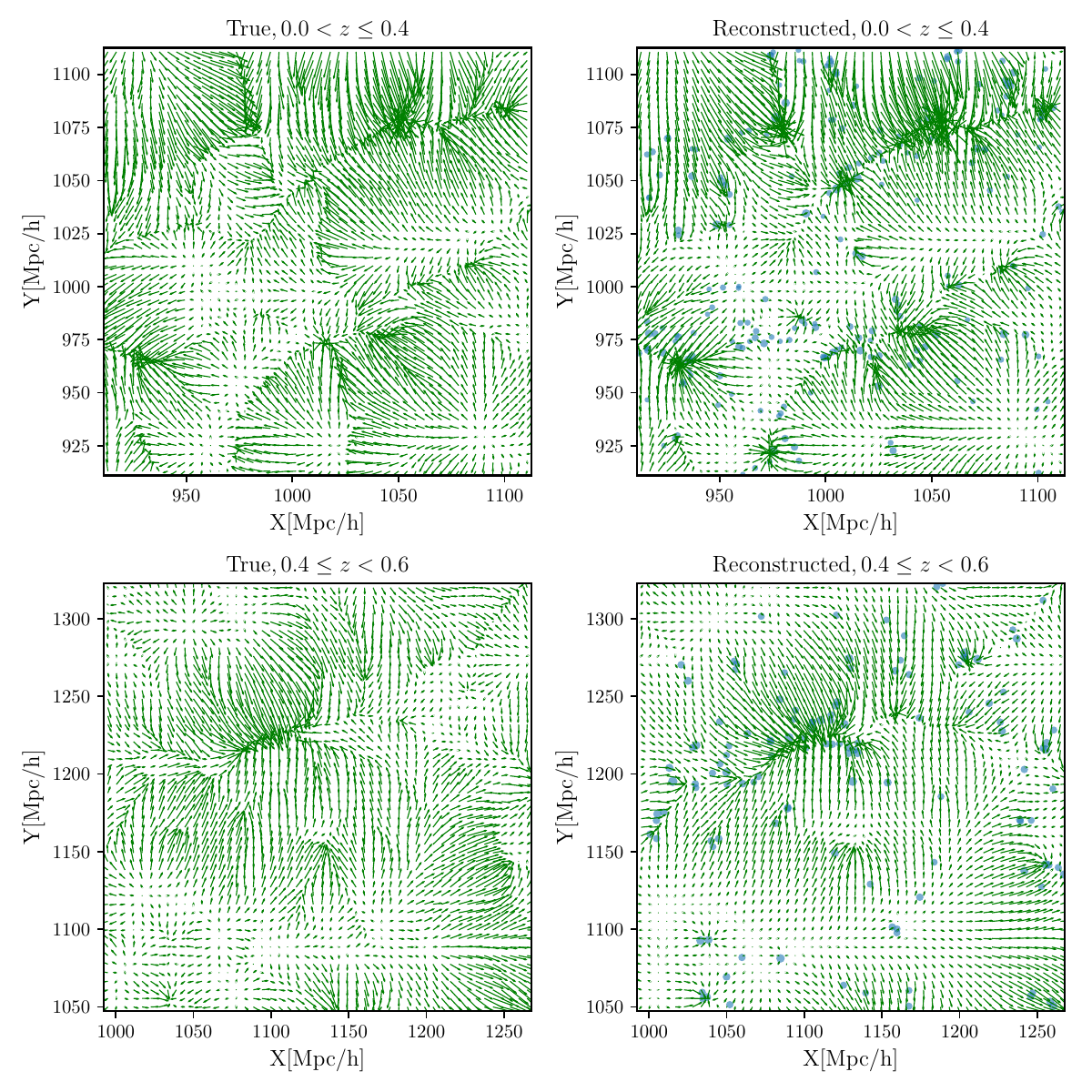}
    \caption{A slice of velocity field on X-Y plane at $0.0 < z \leq 0.41$ (top panels) and $0.39 \leq  z < 0.6$ (bottom panels), with a thickness of 4.03 and 5.51 $\mpch$. The left and right panels indicate the true and reconstructed velocity field, respectively. The blue points represent galaxy groups used in the reconstruction, with symbol sizes proportional to group mass.}
    \label{fig:vel_comp}
\end{figure*}

\begin{figure*}
    \centering
    \includegraphics[width=0.8\linewidth]{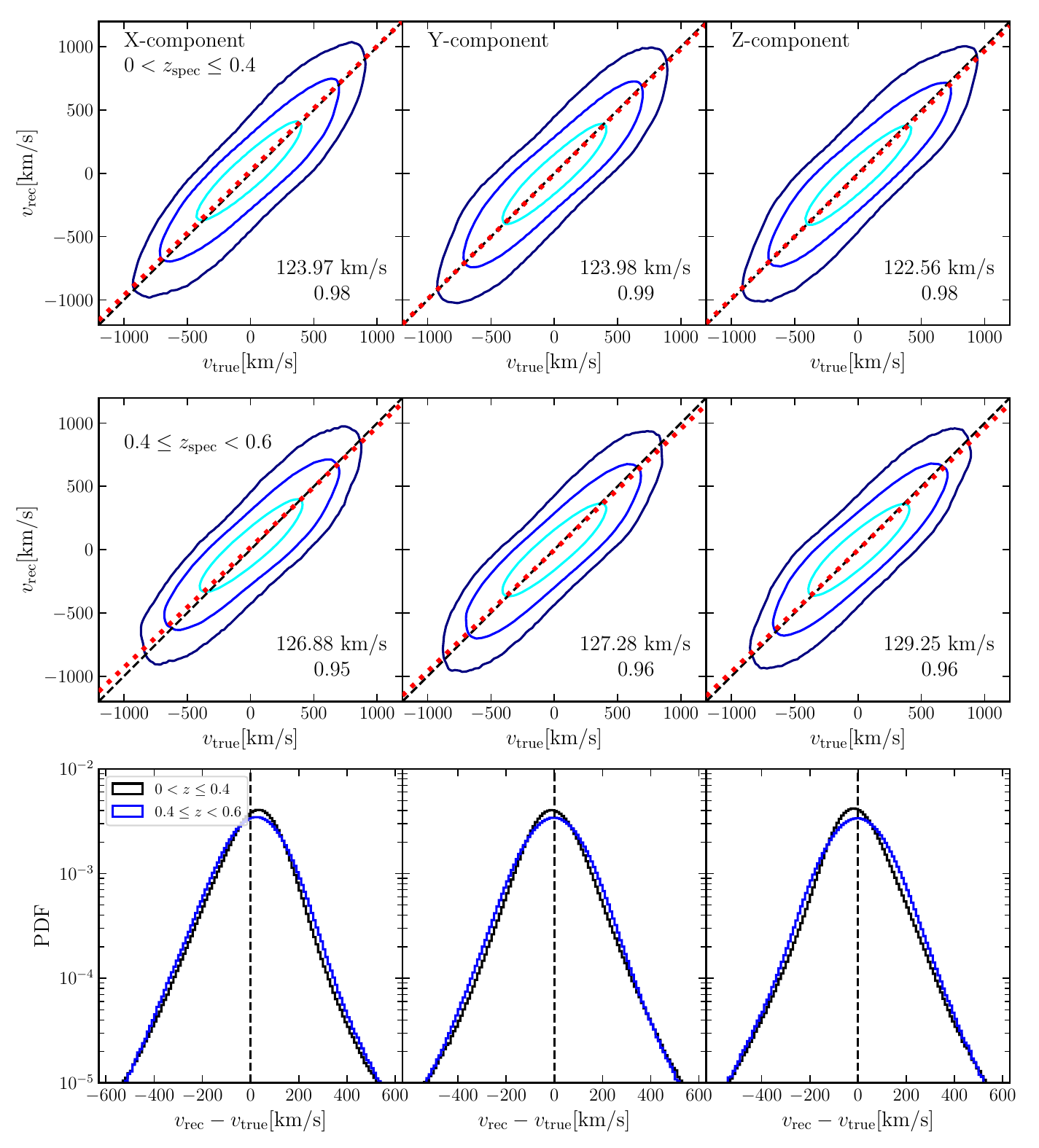}
    \caption{Grid-to-grid comparison between the reconstructed and true velocity fields at $0.0 < z \leq 0.41$ and $0.39 \leq z < 0.6$. The grid cells within $0.0 < z \leq 0.4$ and $0.4 \leq z < 0.6$ are used to evaluate for the two redshift bins. 
    % The dashed and solid lines represent all the grid cells and the grid cells with $0.9 \leq F_{80} \leq 1$ (i.e., excluding boundary grids), respectively. 
    The left, middle, and right sub-panels show the velocity components at $X$-, $Y$-, and $Z$- axis. The contours with cyan, blue, and deep blue colors represent 67, 95, and 99 per cent of the grid cells. The red dotted line shows the best-fitting linear relation. The numbers shown in the right bottom of sub-panels indicate the scatter around the best-fitting line and slope of the best-fitting. The bottom panels show the probability distribution function of the velocity difference in grid cells between the reconstructed and true velocities.}
    \label{fig:vel_specz}
\end{figure*}

\subsection{Tidal field} \label{sec:tidal}

The tidal field describes the gravitational potential through the second derivatives of the gravitational potential, also known as tidal tensor, which is defined as 
\begin{equation} \label{eq:tidal1}
    \mathcal{T}_{ij} = \partial_i \partial_j \phi,
\end{equation}
where $\phi$ is the peculiar gravitational potential.The gravitational potential can alternatively be obtained from the Poisson equation, given by
\begin{equation} \label{eq:tidal2}
    \nabla^2 \phi = 4 \pi G \bar{\rho}a^2\delta_m,
\end{equation}
where $\delta_m$ denotes the dark matter density field. In the practical application, $\phi$ is obtained by solving the Poisson equation in Fourier space. The tidal tensor is an important and effective quantity to describe the rate of spatial change of the gravity field, deciding whether an object is stretched or compressed in a space position. The eigenvalues of tidal tensor quantify the intensity of stretched (or compressed) along different direction of space and are generally used to classify the environments of cosmic web where halos and galaxies reside. The grid cells based on the eigenvalues of tidal tensor $T_1$, $T_2$, and $T_3$ ($T_1$ > $T_2$ > $T_3$), are divided into four types: \emph{cluster}, \emph{filament}, \emph{sheet}, and \emph{void} \citep[e.g.,][]{Hahn2007, ZhangY2024}. Specifically, grid cells with one, two, and three negative eigenvalues are classified as \emph{filament}, \emph{sheet}, and \emph{void}, while grid cells with three positive eigenvalues are regarded as \emph{cluster}. 

In this work, we calculate the tidal field and the eigenvalues of the tidal tensor using the reconstructed mass density field as illustrated in Section.~\ref{sec:method-mass}. The eigenvalues of the tidal tensor are used to classify cosmic web.

% In this section, we evaluate the performance of the reconstructed velocity fields and the effectiveness of the correction of Kaiser effect. 
% We also discuss the effect of survey boundaries and the resolution of the survey box on the reconstructed velocity field.     

\begin{figure*}
    \centering
    \includegraphics[width=0.70\textwidth]{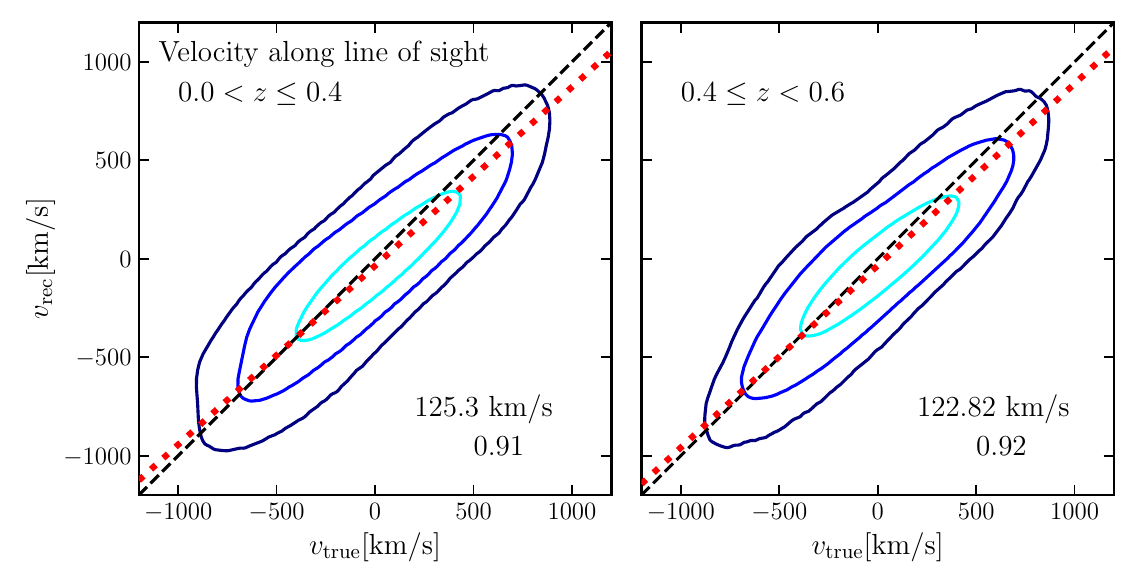}
    \caption{Similar to Figure~\ref{fig:vel_specz} but for the grid velocity along line of sight between the reconstructed and true velocity field in different redshift bins. }
    \label{fig:vel_specz_los}
\end{figure*}

\begin{figure*}
    \centering
    \vspace{-0.5cm}
    \includegraphics[width=0.7\textwidth]{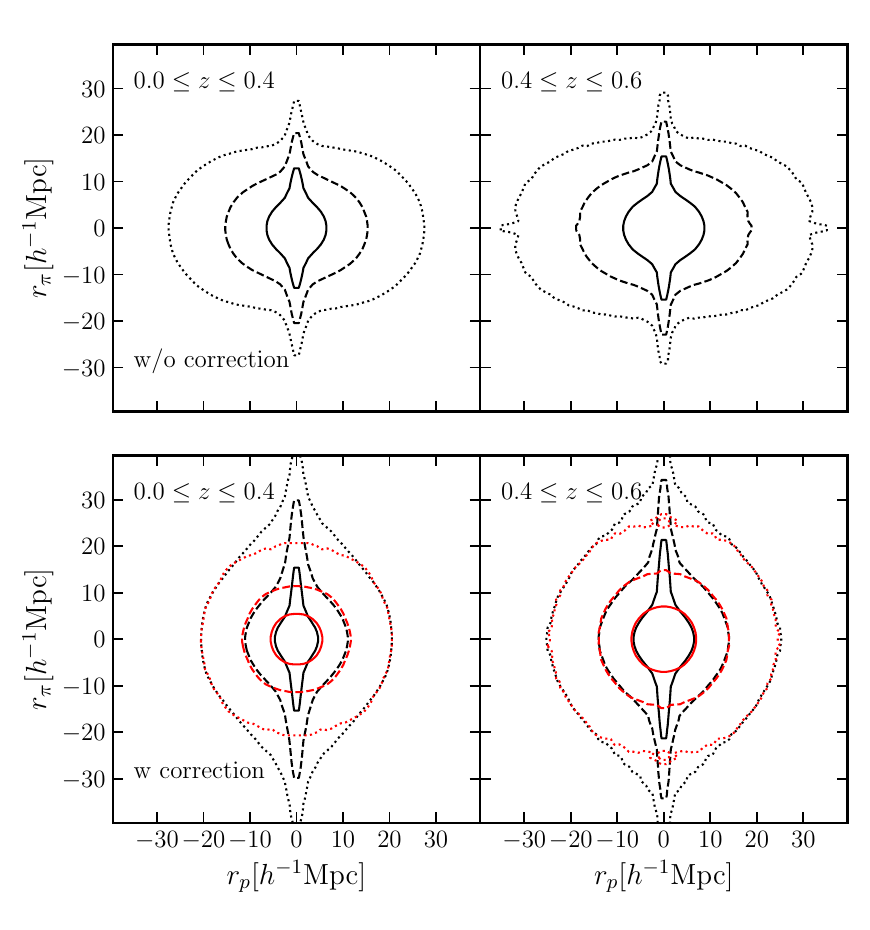}
    \caption{Two-dimensional two-point correlation functions of galaxies. The top and bottom panels present the results without and with correction of Kaiser effect using the reconstructed velocity field, respectively. The left and right columns correspond to different redshift ranges. The solid, dashed, and dotted black contours indicate the level of correlation function $\xi = 1, 0.3, 0.1$. The red lines represent the real-space clustering of galaxies calculated with cosmological redshift from the mock catalog.
    }
    \label{fig:xi2d_specz}
\end{figure*}

\section{Testing the performance using mock galaxy and group catalogs} \label{sec:results}

We employ a galaxy and group catalog that accounts for the primary survey selection effects to assess to what extent the MTV of our universe can be recovered from the DESI DR3 data.

\subsection{Velocity field} 

\subsubsection{Basic performance}

We assess the performance of the reconstructed velocity fields in three aspects: the velocity vector direction, each individual velocity component, and the velocity along the line of sight.

We begin with an intuitive comparison by showing a slice of the velocity vector field, as illustrated in Figure~\ref{fig:vel_comp}. The length of velocity vector is proportional to the magnitude of velocity. At $0.0 < z \leq 0.4$, the reconstructed velocity recover similar velocity patterns as true ones at both dense and sparse group regions. The direction of the velocity reliably follows the spatial distribution of galaxy groups, such that regions containing many galaxy groups exhibit a large velocity magnitude. At $0.4 \leq z < 0.6$, the velocity field in dense regions is well captured, whereas the velocity structure in sparse areas is recovered with lower accuracy. Although the velocity statistics in the two redshift bins are converging (see, e.g., Figure~\ref{fig:vel_comp}), the reconstruction describes the velocity field at higher redshift less accurately. This reduced performance arises from the higher halo mass threshold used in the reconstruction, which leaves low-density regions poorly sampled and thus inadequately characterized. In addition to these qualitative inspections, we also quantitatively show that the reconstructed velocity vectors are well aligned with the true ones by computing the cosine of the angle between their velocities, as described in Appendix~\ref{sec:dir_vel}.

Next, we assess the quality of the reconstructed velocity field by comparing it, cell by cell, with the true velocity for all three velocity components, as illustrated in Figure~\ref{fig:vel_specz}. The true velocity, $v_{\rm true}$, is defined as the average velocity within each grid cell, smoothed with a Gaussian kernel on the same scale as the underlying density field, and calculated using one percent of the dark matter particles. We focus on the grid cells with redshift located at $0.0 < z \leq 0.4$ and $0.4 \leq z < 0.6$ which are used to correct linear RSD effect in practice. Overall, the reconstructed velocity, $v_{\rm rec}$, agrees well with the true velocity, demonstrating that the reconstruction technique remains robust even for extremely large survey volumes.
We quantify the relation between the reconstructed and true velocity field using a linear model, $y(x) = ax+b$, where $a$ and $b$ are free parameters representing the slope and intercept, respectively. The best-fitting parameters are determined using orthogonal distance regression (ODR), which minimizes the orthogonal distances of the data points from the fitted relation, rather than the residuals along the $y$-direction alone. The linear fitting using the ODR algorithm is implemented in the \texttt{scipy.odr} module of \texttt{SciPy} \citep{SciPy2020}.
The best-fitting linear relations between $v_{\rm rec}$ and $v_{\rm true}$, shown as red dotted lines, further demonstrate the tight correlation between the two velocity fields. The slopes of these relations are close to unity at both redshift bins, confirming that the reconstruction is largely unbiased. 
At $0.0 < z \leq 0.4$, the typical velocity residuals measured around the best-fitting relation is $\sim 123 \kms$. At higher redshift bin, the scatter increases mildly, reaching $\sim 130 \kms$. 
% The increase of scatter can be attributed to the increasing group mass threshold at higher redshift, which leads to a sparser sampling of the underlying density field, thereby reducing the accuracy of the reconstructed velocity field. 
Further, we show the probability distribution of the velocity difference between the reconstructed and true velocities in the bottom panels of Figure~\ref{fig:vel_specz}. The velocity difference is almost unbiased at three velocity components in the two redshift bins.

\begin{figure*}
    \centering
    \includegraphics[width=0.85\textwidth]{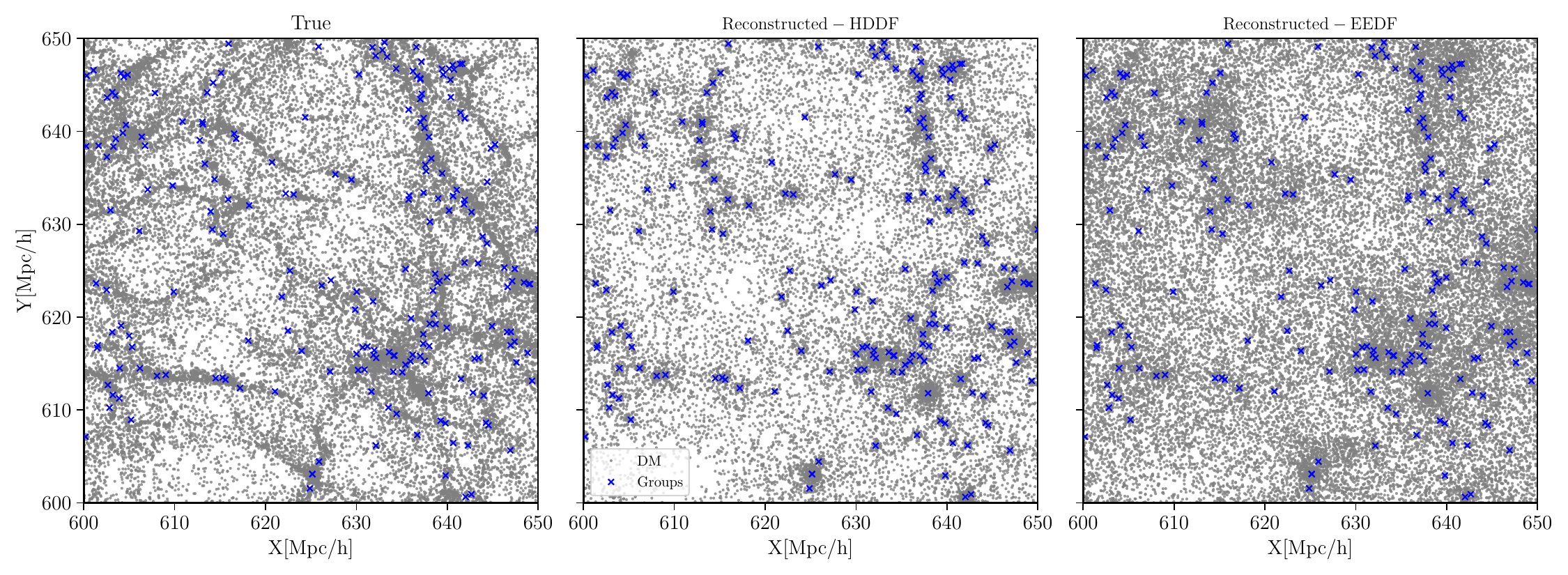}
    \caption{A Comparison of the DM particle distribution in a slice of the X-Y plane with a thickness of 10 $\mpch$, extracted from the true density field (left panel) and reconstructed HDDF (middle panel) and EEDF (right panel) at $0.0 < z < 0.4$. The grey points represent DM particles, while the blue crosses show the position of the galaxy groups with the Kaiser effect corrected. Only one per cent of DM particles in the true density field is shown here. We here hold the DM particle number density in EEDF same as true density field.}
    \label{fig:dens_pt}
\end{figure*}

\begin{figure*}
    \centering
    \includegraphics[width=0.9\textwidth]{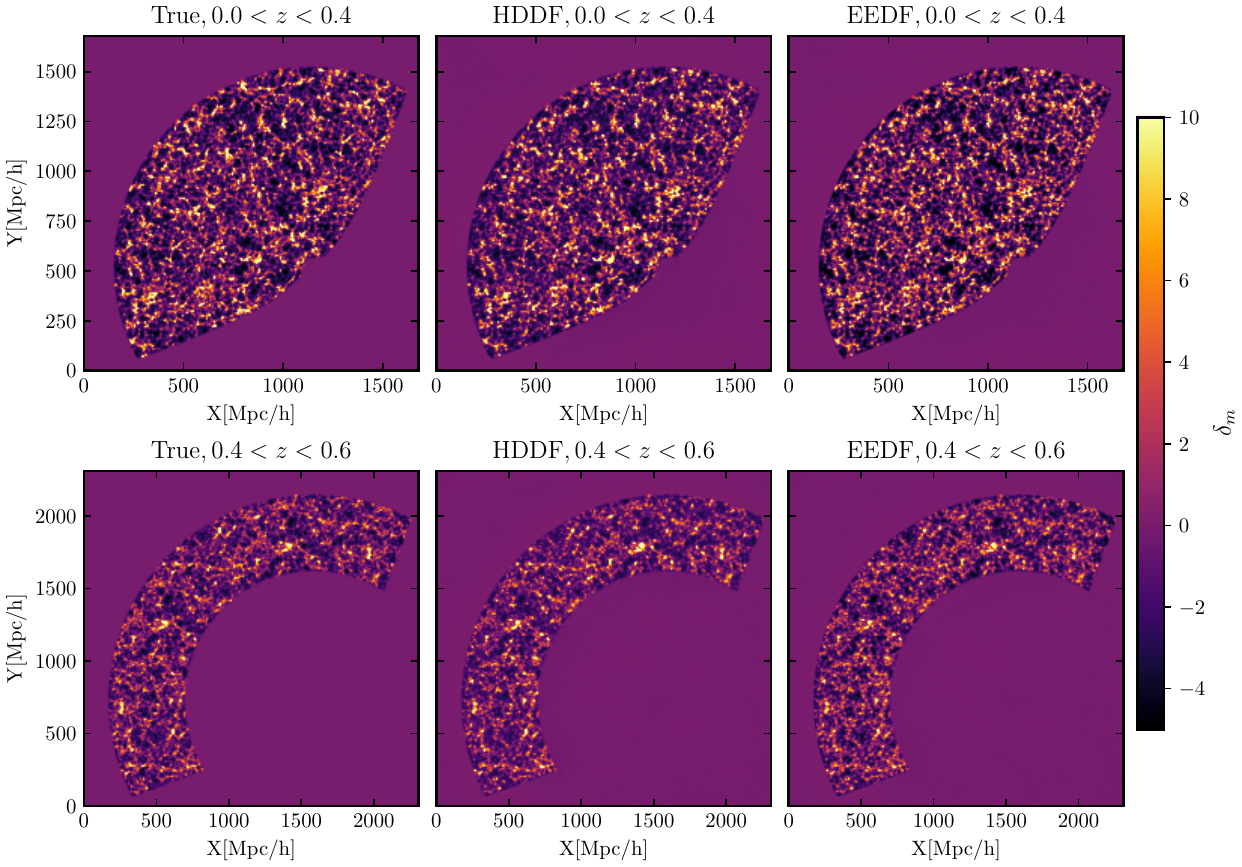}
    \caption{A comparison of density slice between the true density field (left panel) and reconstructed HDDF (middle panel) and EEDF (right panel). The top and bottom panels show the mass density fields at $0.0 < z < 0.4$ and $0.4 < z < 0.6$ with a thickness of 24.2 and $33.1 \mpch$ approximately.}
    \label{fig:dens_comp}
\end{figure*}

In observation, the velocity along the line of sight, $v_{\rm los}$, contains significant cosmological information and is used to correct the RSD effect. Therefore, we inspect the performance of $v_{\rm los}$ for grid cells as shown in Figure~\ref{fig:vel_specz_los}. In general, the $v_{\rm los}$ between the reconstructed and true values is consistent with each other, showing a typical error of $\sim 120 \kms$. The slope of the best-fitting in the two redshift bins is approximately 0.91, which is slightly lower compared to the single velocity component in Figure~\ref{fig:vel_specz}. As the calculation of $v_{\rm los}$ depends on the three components of the velocity, the errors of $v_{\rm los}$ are also contributed by the three components of the velocity.

\subsubsection{Application: correction of Kaiser effect}

An important use of the velocity field is to account for redshift-space distortions (RSD), especially the Kaiser effect. Therefore, we assess how well the reconstructed velocity field mitigates the Kaiser effect by analyzing the two-dimensional two-point correlation function (2PCF).

We compute 2PCF using a galaxy sample selected from the galaxy group catalog. The reconstructed velocity field is used to correct the redshift according to Equation~\ref{eq:zcorr}. The two-dimensional 2PCF corrected for the Kaiser effect and without is presented in Figure~\ref{fig:xi2d_specz}. The solid, dashed, and dotted contours correspond to correlation levels of $\xi = 1$, $0.3$, and $0.1$, respectively. We also included a 2PCF computed from cosmological redshift, i.e., without the RSD effect, as a reference (red lines in Figure~\ref{fig:xi2d_specz}). After the redshift correction, the Kaiser effect is significantly eliminated in both redshift bins as indicated by the circular shape of the different levels of 2PCF. However, the FoG effect becomes obvious after the correction of Kaiser effect as shown in the lengthen shape along line of sight. We will consider eliminating the FoG effect in future work. Overall, the performance of 2PCF indicates that the reconstructed velocity field is effective for the correction of the Kaiser effect in the two redshift bins.

\begin{figure*}
    \centering
    \includegraphics[width=0.7\textwidth]{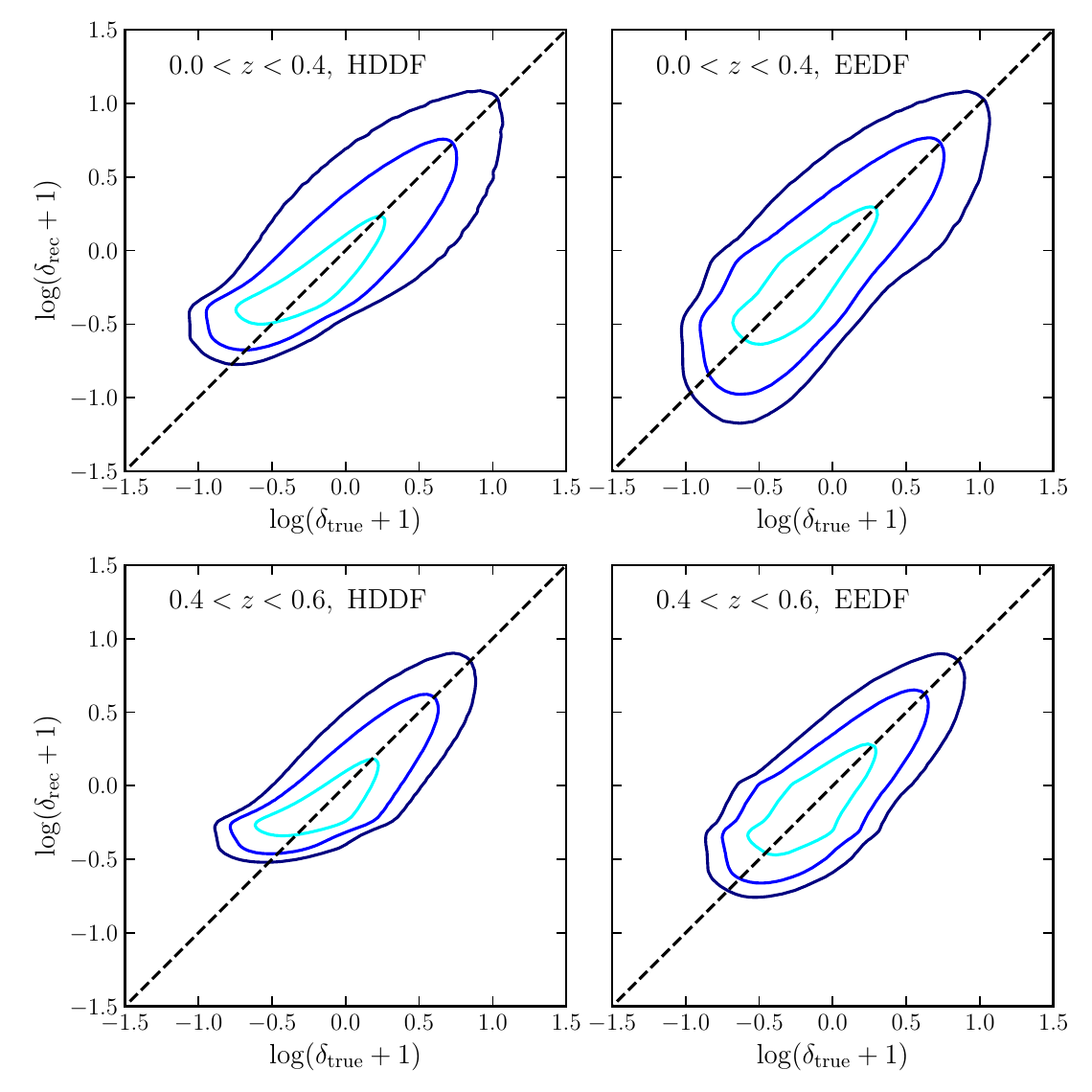}
    \caption{Grid-to-grid comparison of density value between the reconstructed and true fields. The top and bottom panels show the comparisons at $0.0 < z < 0.4$ and $0.4 < z < 0.6$, while the left and right panels represent reconstructed HDDF and EEDF, respectively. The contours with cyan, blue, and deep blue colors represent 67, 95, and 99 per cent of the grid cells.}
    \label{fig:delta}
\end{figure*}

\subsection{Mass density field} \label{sec:massdens}

In this subsection, we assess the performance of the reconstructed mass density field. We compare it with the true mass density field, examining the large-scale dark matter distribution and the values of the density contrast field. We also present reconstruction results at different resolutions of the survey box.

\subsubsection{Basic performance}

We start by investigating the small-scale density fields by focusing on a subregion of the survey volume within $0.0 < z < 0.4$, as shown in Figure~\ref{fig:dens_pt}, before applying the phase-space abundance matching method. In this area, we display the positions of galaxy groups along with the sampled DM particles used to build the density fields. The left panel shows the distribution of the true dark matter particles. After correcting the galaxy group positions for the Kaiser effect and placing them in the true field, most groups align closely with the actual DM distribution, although a few still exhibit slight positional offsets. The DM particles generated by HDDF follow the spatial arrangement of the galaxy groups closely and clearly separate cluster regions from voids in the middle panel of Figure~\ref{fig:dens_pt}. These sampled DM particles reproduce the overall structure accurately. When we include the environmental background DM particles, the contrast between different environments diminishes, but the depiction of regions outside galaxy groups becomes more accurate. Overall, we observe that the sampled DM particles still follow the high-density regions and primary filamentary structures in a way that remains consistent with the underlying true DM distribution.

Further, we assess how effectively the reconstructed density fields capture large-scale structures after applying the phase-space abundance matching technique. In Figure~\ref{fig:dens_comp}, we show a slice-by-slice comparison between the reconstructed and true mass density fields in the two redshift intervals. The true mass density fields are derived from one per cent of the DM particles in a single snapshot of Jiutian-1G, where the snapshot redshift is matched to $z_{\rm eff}$. Overall, both HDDF and EEDF recover a distribution that closely follows the true field in both redshift ranges. The locations of overdense and underdense regions are in good agreement between the reconstructed and true density fields. It is, in fact, challenging to visually distinguish the reconstructed density fields from the true one. Taken together, the global density distribution in this snapshot indicates that our method is both promising and robust on large scales.

Moreover, in Figure~\ref{fig:delta} we compare, cell by cell, the density contrast of the reconstructed field with that of the true density field. The values of $\delta$ agree well, especially in high-density regions, for both HDDF and EEDF across the two redshift intervals. This is anticipated because the halos used in the halo-domain sampling scheme are predominantly situated in overdense environments. The halo-domain sampling method also enables us to sample DM particles in underdense regions, particularly inside cosmic voids, by choosing a sufficiently large sampling radius. Including DM particles well outside the halo boundaries improves our description of low-density regions. However, the paucity of galaxy groups in these areas limits the reconstruction fidelity and introduces a mild systematic bias in the recovered density field. This effect becomes more pronounced in the $0.4 < z < 0.6$ bin. The environment-enhanced sampling method mitigates this shortcoming by exploiting information from the large-scale environment, reducing the bias at low-density regions, although it also adds extra noise, resulting in a slightly larger scatter. We have also confirmed the better performance of EEDF by performing a quadratic fit, whose best-fitting parameter provides a measure of how strongly the data deviate from a linear correlation \citep[e.g.,][]{Fang2024}. A more comprehensive comparison of the HDDF and EEDF performance in reconstructing both the initial conditions and the final density field will be presented in a forthcoming ELUCID-DESI III paper.

\subsubsection{Resolution of survey box}

\begin{figure}
    \centering
    \includegraphics[width=0.45\textwidth]{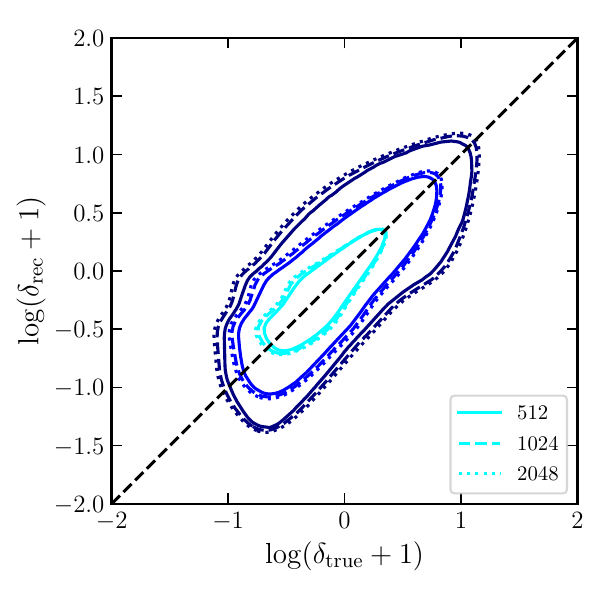}
    \caption{Comparison of reconstructed EEDF in different resolutions of survey box at $0.0 < z < 0.4$. The solid, dashed, and dotted lines represent the density field reconstructed with $512^3$, $1024^3$, and $2048^3$ grids, respectively. The contours with cyan, blue, and deep blue colors represent 67, 95, and 99 per cent of the grid cells.}
    \label{fig:dens_resol}
\end{figure}

In this subsubsection, we examine how the grid resolution of the survey volume, i.e., the total number of grid cells, affects the reconstructed density fields. We test three resolutions with $512^3$, $1024^3$, and $2048^3$ grid cells. For each case, all generated DM particles are first deposited onto the grid using the CIC scheme, and then subjected to Gaussian smoothing with an identical smoothing scale. In Figure~\ref{fig:dens_resol}, we compare the reconstructed EEDF densities with the true density field in the redshift range $0.0 < z < 0.4$ for these different resolutions. The contours exhibit almost the same shapes and occupy nearly the same locations, demonstrating that the reconstructed density field is largely insensitive to the grid resolution over the range explored. This finding justifies the use of a higher grid resolution, which is crucial for recovering the initial conditions in our ELUCID-DESI project.

\subsection{Tidal field} \label{sec:tidal}

The tidal fields are computed from the reconstructed matter density field using Eqs.~\ref{eq:tidal1} and \ref{eq:tidal2}. We denote the tidal fields reconstructed from HDDF and EEDF as the Halo-Domain Tidal Field (HDTF) and the Enhanced-Environment Tidal Field (EETF), respectively. In this subsection, we assess the accuracy of the reconstructed tidal field and use its characteristics to classify the large-scale environments. 

\begin{figure*}
    \centering
    \includegraphics[width=0.85\linewidth]{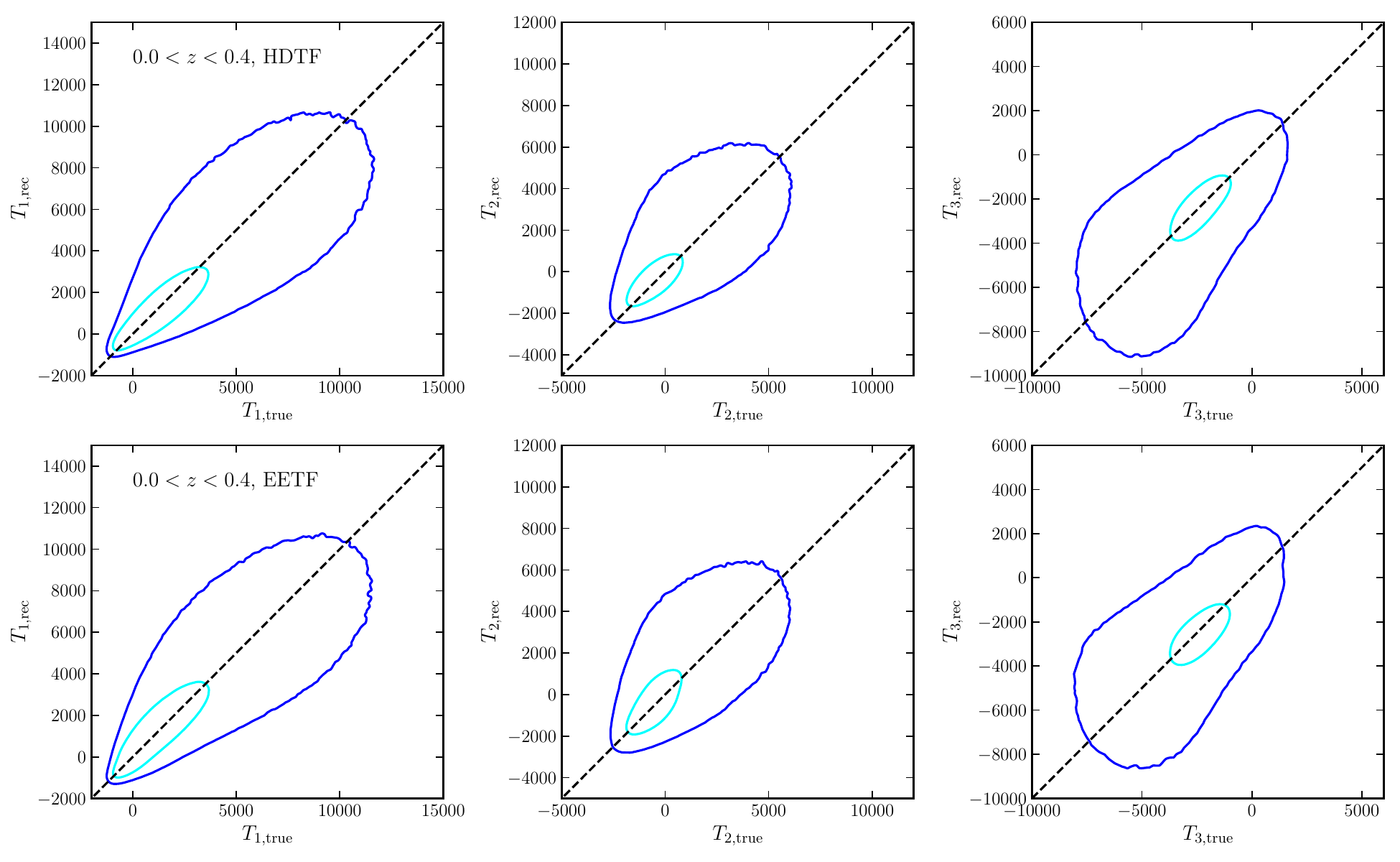}
    \caption{Comparison of the eigenvalue of tidal tensor between the reconstructed and true tidal field at $0.0 < z < 0.4$. The left, middle, and right panels indicate the comparison for the different components of eigenvalue of tidal tensor. The top and bottom panels show the comparison of HDTF and EETF, respectively. The contours with cyan and blue colors represent 67 and 95 per cent of the grid cells.}
    \label{fig:tidal_eigenv_low}
\end{figure*}

\subsubsection{Eigenvalue comparison}

\begin{figure*}
    \centering
    \includegraphics[width=0.85\linewidth]{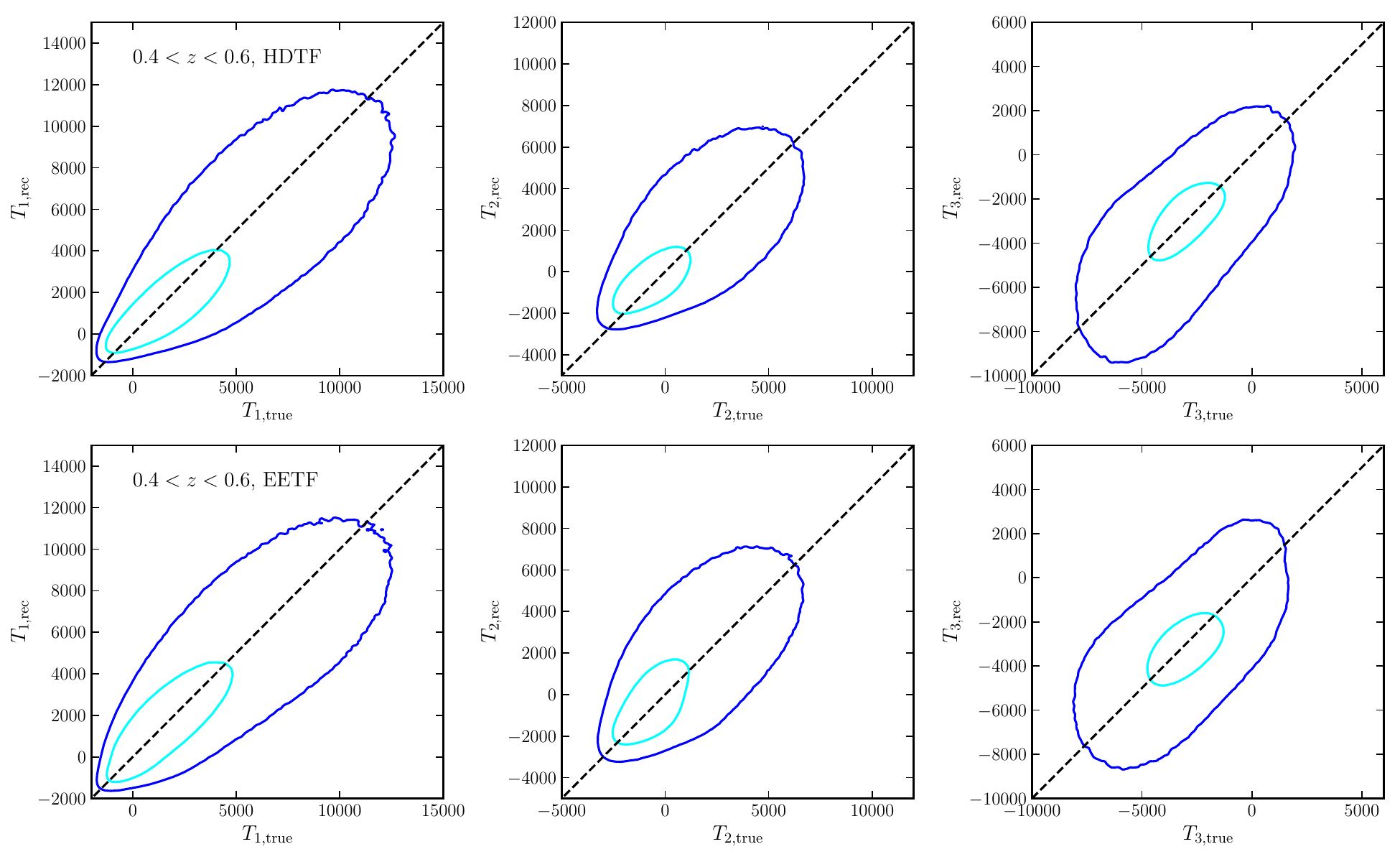}
    \caption{Similar as Figure~\ref{fig:tidal_eigenv_low} but for the redshift range $0.4 < z < 0.6$}
    \label{fig:tidal_eigenv_high}
\end{figure*}

The reliability of the reconstructed tidal field can be evaluated using the eigenvalues of the tidal tensor. In Figure~\ref{fig:tidal_eigenv_low}, we compare the three eigenvalues of the tidal tensor from the reconstructed HDTF and EETF at $0.0 < z < 0.4$ with those derived from the true tidal field. Each eigenvalue largely aligns along the diagonal, indicating strong consistency between the reconstructed and true tidal fields. The contours enclosing 95 per cent of the grid cells (blue contours) further show that the reconstructed eigenvalues track the true ones closely, confirming the fidelity of the tidal field reconstruction. In addition, HDTF and EETF share similar percentile contours, implying that the inclusion of environment-enhanced dark matter particles does not substantially impact the tidal field reconstruction. The results for the redshift range $0.4 < z < 0.6$, shown in Figure~\ref{fig:tidal_eigenv_high}, display performance comparable to that in the $0.0 < z < 0.4$ bin.

\begin{figure*}
    \centering
    \includegraphics[width=0.85\linewidth]{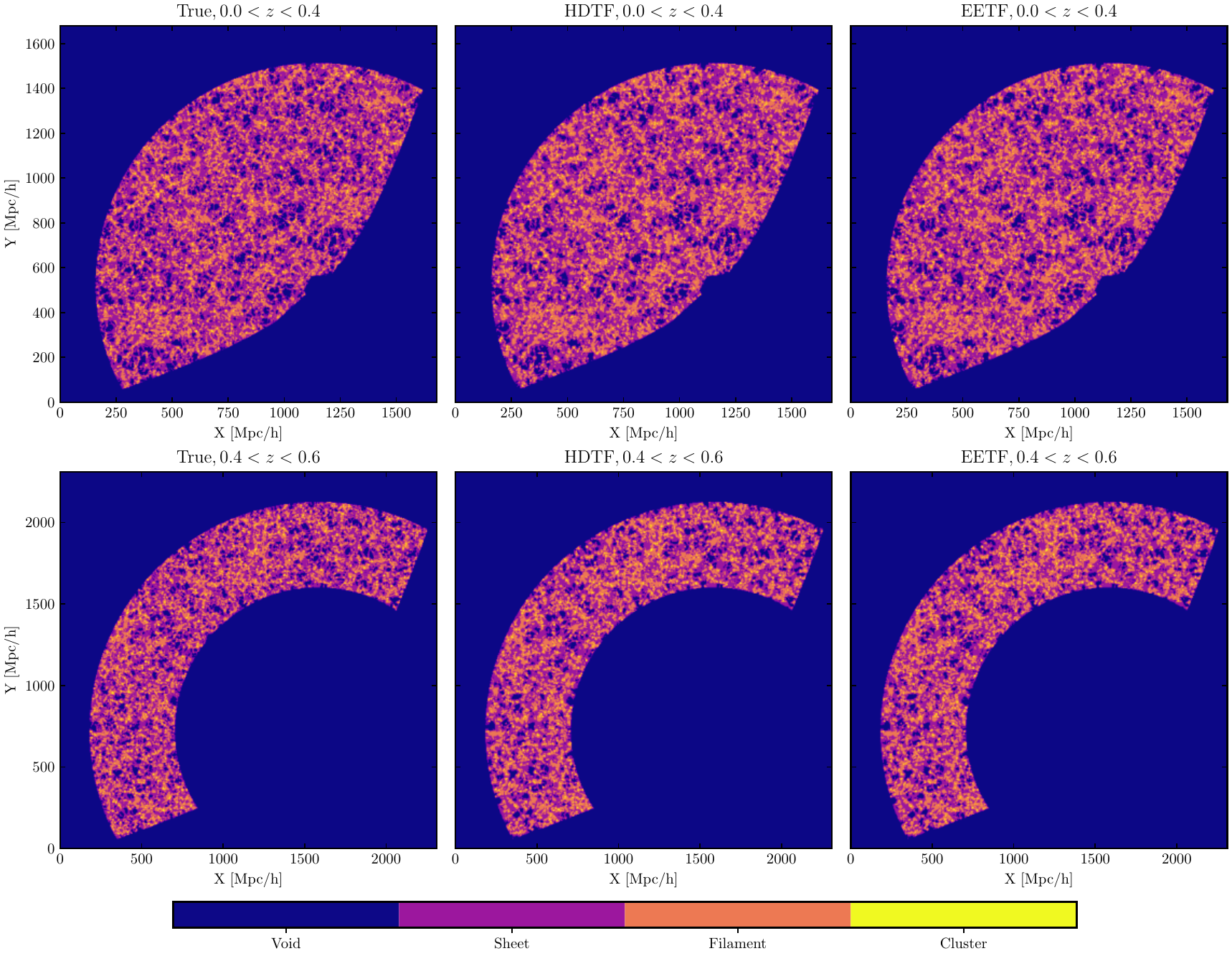}
    \caption{A Comparison of the classification of large-scale environments for a slice of tidal field at $0.0 < z < 0.4$ (top panel) and $0.4 < z < 0.6$ (bottom panel). The cluster, filament, sheet and void are expressed with yellow, orange, magenta, and blue colors. The left, middle, and right panels represent the classification for the true tidal field, HDTF, and EETF, respectively.}
    \label{fig:env}
\end{figure*}

\begin{figure*}
    \centering
    \includegraphics[width=0.85\textwidth]{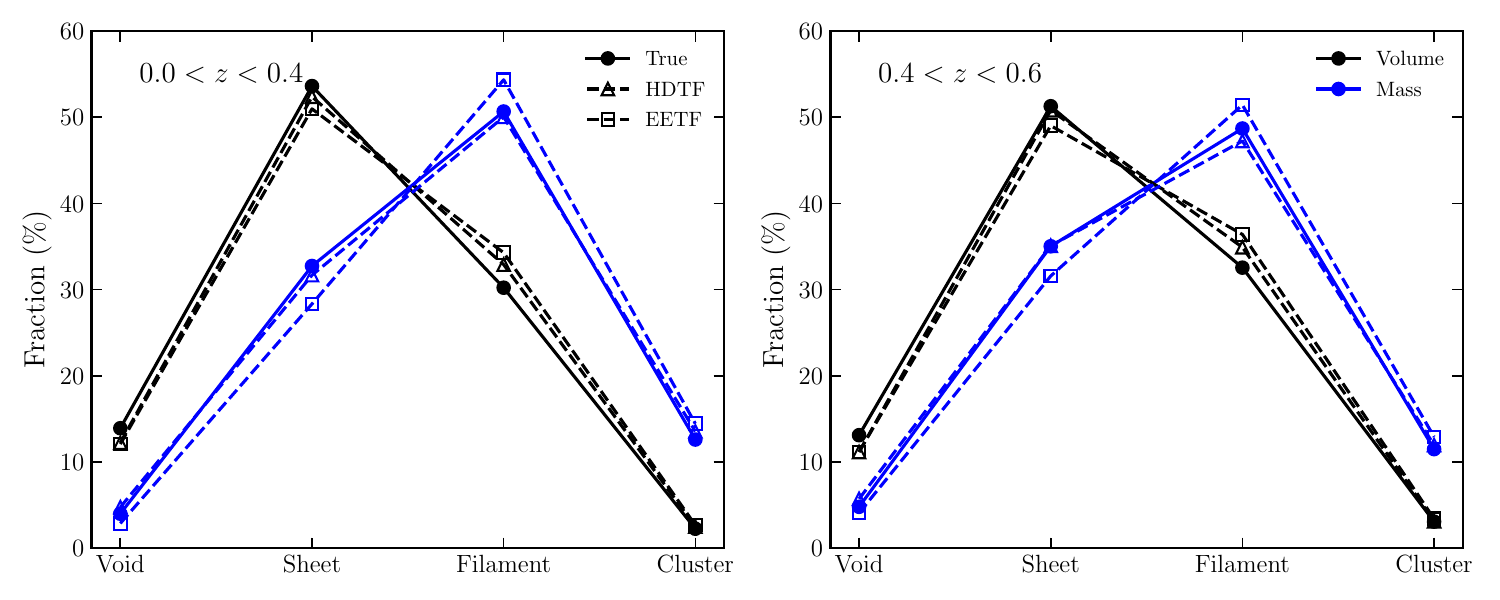}
    \caption{Fractions of different large-scale environments for the tidal fields at $0.0 < z < 0.4$ (left panel) and $0.4 < z < 0.6$ (right panel). The black and blue colors indicate the volume fraction and mass fraction in grid cells. The circles, triangles, and squares represent the fractions from the true, reconstructed HDTF and EETF, respectively.}
    \label{fig:stat_tidal}
\end{figure*}

\subsubsection{Classification of Environment}

The tidal tensor characterizes how gravity stretches or compresses matter along different directions. It is therefore widely employed to classify the large-scale environment, a procedure commonly referred to as the T-Web method \citep[e.g.,][]{Hahn2007,Forero2009,Aycoberry2024}.  
In this framework, the components of the cosmic web are categorized as \emph{cluster}, \emph{filament}, \emph{sheet}, or \emph{void}, according to how many eigenvalues of the tidal tensor in each grid cell are positive.  
Figure~\ref{fig:env} presents the distribution of these large-scale environments in a slice of the reconstructed fields. The environmental classifications obtained from the reconstructed tidal fields closely match those derived from the true tidal field in both redshift intervals. No notable discrepancies are observed between the reconstructions from HDTF and EETF. Furthermore, the inferred environmental distribution aligns with theoretical expectations; for instance, clusters tend to appear at the junctions of filaments. Overall, the reconstructed tidal fields yield a reliable and robust classification of the large-scale cosmic environments.

Furthermore, in Figure~\ref{fig:stat_tidal} we present the volume and mass fractions associated with the different cosmic web environments. Among the four types, filaments occupy the largest share of the volume, whereas clusters occupy the smallest, in line with previous T-web analyses based on cosmological simulations \citep[e.g.,][]{Forero2009,Cui2018}. In contrast, filaments and voids contain the highest and lowest mass fractions, respectively, for all considered density fields, consistent with \citet{Forero2009} and \citet{Libeskind2018}. Since the exact numerical values of these fractions depend on several factors \citep[see][]{Forero2009, Libeskind2018}, and our primary aim here is to validate the field reconstruction method, we emphasize the agreement of environmental fractions across different tidal fields rather than their absolute values.
The reconstructed tidal fields reproduce the cluster volume fraction very accurately relative to the true tidal field, but they slightly underestimate the volume fractions of void and sheet environments. 
The mass fractions obtained from HDTF more closely match those from the true tidal field, especially for sheets and filaments.
In addition, neither the volume nor the mass fractions exhibit noticeable evolution across the examined redshift intervals.
Taken together, the environmental fractions derived from HDTF are highly consistent with those from the true tidal field, indicating that incorporating environment-enhanced dark matter particles has only a minimal effect on the resulting cosmic web classification.

\section{Conclusions} \label{sec:summary}

We present a new approach to reconstruct the dark matter mass density, tidal, and velocity (MTV) fields using galaxy groups. Rather than relying on the theoretical bias correction, our method connects galaxy groups to the underlying dark matter density field through their phase-space information. This avoids a major source of systematic error and enables a direct application to spectroscopic redshift surveys. We test and assess the performance of the method using a DESI light-cone mock catalog that includes realistic observational selection effects. The galaxy group catalog is generated with a halo-based group finder applied to DESI mock galaxies limited to an apparent magnitude of $m_z < 19.65$ over $0.0 < z < 0.6$, yielding a sample comparable in size to the DESI BGS faint sample ($m_r < 20.175$). The reconstruction pipeline begins by inferring the velocity field to correct the Kaiser effect in the group distribution, and then reconstructs the mass density field from this corrected group catalog. Dark matter particles are populated following two schemes. In the first, particles are assigned based on the positions and masses of the galaxy groups using the halo-domain method. In the second, additional particles are added to capture the large-scale background environment. From the corresponding reconstructed density fields, HDDF and EEDF, we derive the tidal field and use it to classify the large-scale environment. Overall, we reconstruct the MTV fields in two redshift ranges, $0 < z < 0.4$ and $0.4 < z < 0.6$. Our main results can be summarized as follows:

\begin{itemize}
    \item The reconstructed velocity field shows no bias in the grid-to-grid comparison for all three velocity components, exhibiting a typical scatter of $\sim 120\ \rm \kms$ and $130\ \kms$ in the ranges $0.0 < z < 0.4$ and $0.4 < z < 0.6$, respectively. The velocity vector map clearly illustrates the matter flow over the full redshift interval, although regions at higher redshift are more sparsely characterized because of the stringent group mass limit.

    \item The reconstructed velocity field can be effectively employed to correct the Kaiser effect throughout the full redshift range probed by the two-dimensional 2PCF.

    \item The reconstructed HDDF and EEDF closely resemble the true density field, maintaining a consistent DM distribution in both high- and low-density regions. The EEDF exhibits a slightly higher density amplitude, attributable to the inclusion of environmental background particles. %The power spectra of the reconstructed and true density fields agree very well up to $k \sim 0.4$ in both redshift bins.

    \item The grid density values obtained with HDDF closely match those of the true density fields in high-density regions, but exhibit a moderate discrepancy in low-density areas, where the number of groups is small. By integrating large-scale environmental information, EEDF further improves the reconstruction of low-density structures, highlighting its promise for enhancing mass field reconstruction in future applications.  

    \item Within the tested range, the reconstructed density fields show no significant dependence on the chosen grid resolution, which supports employing high-resolution grids for future initial-condition reconstructions in the ELUCID-DESI project.

    \item The reconstructed HDTF and EETF are in good agreement with the actual tidal field, successfully recovering the eigenvalues of the tidal tensor over various redshift intervals. The comparable performance of HDTF and EETF suggests that adding environment-enhanced dark matter particles has only a minor effect on the accuracy of the tidal field reconstruction.

    \item The reconstructed tidal fields likewise reproduce the large-scale environmental classifications given by the true tidal field at matching locations in the two redshift bins. The volume fractions of the various environments inferred from HDTF and EETF are in close agreement, differing only by a slightly reduced filament fraction relative to the true field. This indicates that incorporating large-scale environmental information has a minimal effect on the resulting cosmic web classification.
    
\end{itemize}

Overall, our findings show that the group-based phase-space reconstruction technique is a promising and robust method for recovering the dark matter MTV fields. The successful validation with the DESI mock catalog opens the door to its straightforward application to DESI BGS data, which we will explore in future work. For the reconstructed velocity field, the strong performance in the range $0.4 < z < 0.6$ suggests that our approach has considerable potential for use with higher-redshift samples. While the Kaiser effect has been largely removed, the FoG effect is still present and will be tackled in subsequent studies. Furthermore, the reconstructed mass density field provides an important input for setting up initial conditions in the upcoming ELUCID-DESI simulations.

%% Please use the acknowledgment and contribution environments. This will 
%% be anonomyized when the "anonymous" style option is used. 
\begin{acknowledgments}
This work is supported by the National Key R\&D Program of China (2023YFA1607800, 2023YFA1607801, 2023YFA1607804), the National Natural Science Foundation of China (Grant No. 12273088, 12595312, 12595311, 12133006, 12573007, 12125301, 12192222),  “the Fundamental Research Funds for the Central Universities”, 111 project No. B20019, and Shanghai Natural Science Foundation, grant No.19ZR1466800. We acknowledge the science research grants from the China Manned Space Project with Nos. CMS-CSST-2021-A02 \& CMS-CSST-2025-A04. This project is also supported in part by Office of Science and Technology, Shanghai Municipal Government (grant Nos. 24DX1400100, ZJ2023-ZD-001). H.W. and Y.P. acknowledge support from the New Cornerstone Science Foundation through the XPLORER PRIZE. W.C. gratefully thanks Comunidad de Madrid for the Atracci\'{o}n de Talento fellowship no. 2020-T1/TIC19882 and Agencia Estatal de Investigación (AEI, Spain) for the Consolidación Investigadora Grant CNS2024-154838. He is further supported by the Agencia Estatal de Investigación (AEI, Spain) under project PID2024-156100NB-C21, funded by MCIN/AEI/10.13039/501100011033, and ERC: HORIZON-TMA-MSCA-SE under the LACEGAL-III (Latin American-Chinese-European Galaxy Formation Network) project with grant number 101086388 and the science research grants from the China Manned Space Project. Q.L. acknowledges the support from Shanghai Post-doctoral Excellence Program (2025363). 
This work has made use of the Gravity Supercomputer at the Department of Astronomy, Shanghai Jiao Tong University, and is supported by the Kunshan Computing Center.
\end{acknowledgments}

\appendix

\section{Phase consistency} \label{sec:phase}

\begin{figure}
    \centering
    \includegraphics[width=0.48\textwidth]{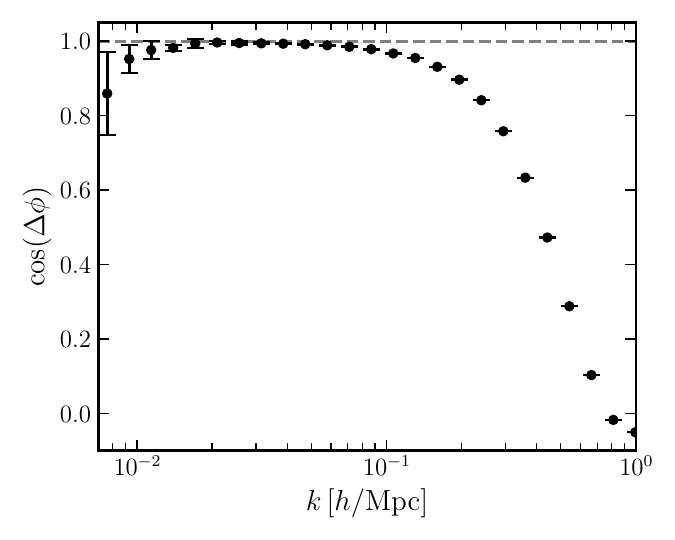}
    \caption{Difference of phase in Fourier space between the galaxy group mass density field and the true mass density field at $0.0 < z < 0.41$. The black points show the medians with error bar representing standard errors.}
    \label{fig:phase_angle}
\end{figure}

\begin{figure*}
    \centering
    \includegraphics[width=0.45\textwidth]{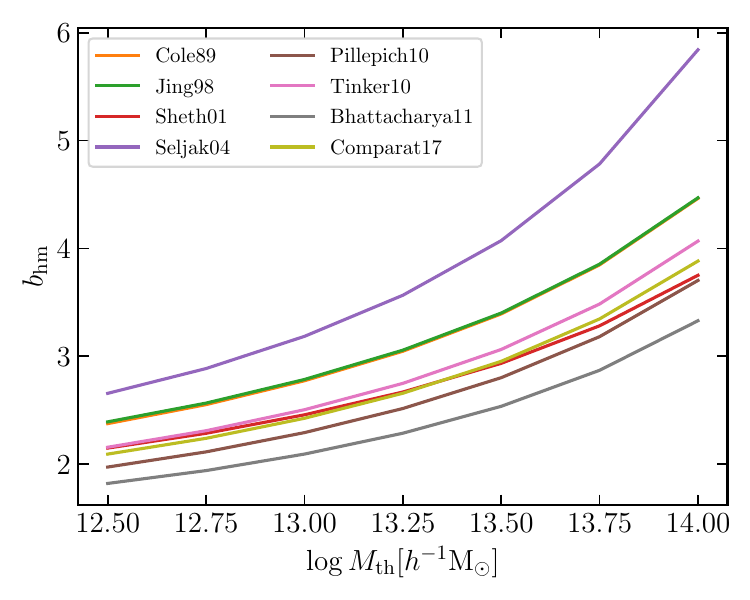}
    \includegraphics[width=0.45\textwidth]{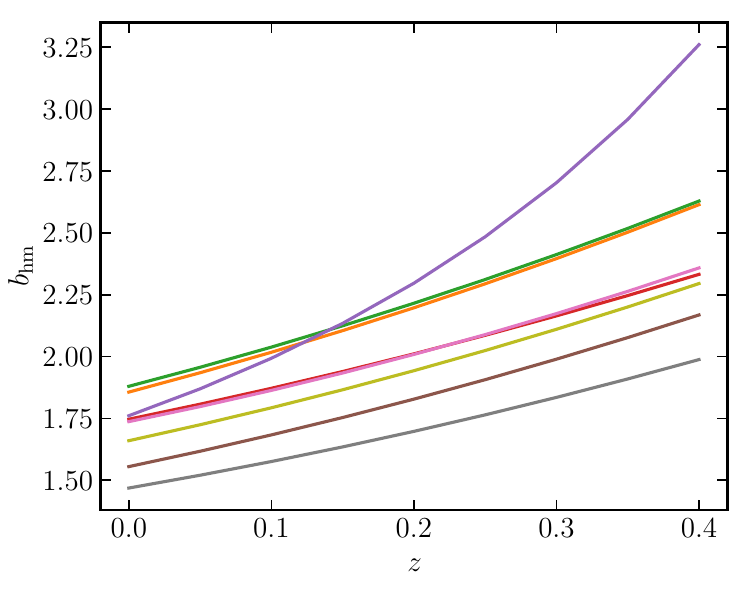}
    \caption{Comparison of the mass-weighted bias parameter predicted from theoretical models. The bias parameter is derived based on a sample of mock groups at $0.0 < z < 0.4$. The lines with different colors represent the bias parameters derived from different theoretical models, including \citet{Cole1989}, \citet{Jing1998}, \citet{Sheth2001}, \citet{Seljak2004}, \citet{Pillepich2010}, \citet{Tinker2010}, \citet{Bhattacharya2011}, and \citet{Comparat2017}. The left panel shows the distribution of mass-weighted bias parameter as a function of halo mass limitation. The redshift used to determine halo bias from theoretical models is fixed to 0.29. The right panel shows the distribution of bias parameter as a function of redshift with mass limitation $M_{th} = 10^{12.5}\Msunh$.}
    \label{fig:biasf_model}
\end{figure*}

In this section, we investigate the difference of phase in Fourier space between the reconstructed galaxy group density field and the true density field, based on the cosine of the phase differences, $\mathrm{cos}(\Delta \phi)$. The phase difference is defined as $\Delta \phi = \phi_{\rm rec} - \phi_{\rm true}$, where $\phi_{\rm rec}$ and $\phi_{\rm true}$ are the phase of galaxy group density field and true density field in Fourier space, respectively. 
In Figure~\ref{fig:phase_angle}, we show the phase difference between the galaxy group density field and the true density field as a function of $k$. As expected, the phases remain highly consistent on large scales ($k \lesssim 0.1\ h \rm Mpc^{-1}$), which supports our use of phase-space abundance matching. Interestingly, the scale at which this consistency begins to break down ($k \sim 0.1\ h \rm Mpc^{-1}$) closely matches the behavior of $b_{hm}$ shown in Fig.~\ref{fig:bhm}. 
Because the Fourier phase carries the positional information of galaxy groups, the phase difference could, in principle, serve as a diagnostic for the effectiveness of RSD corrections that rely on the reconstructed velocity field. However, we do not observe a significant change in the phase differences between the density fields before and after applying the Kaiser correction. This is likely due to the smoothing imposed on the density fields, which may erase the signal associated with the Kaiser-effect correction.

\section{Theoretical mass-weighted bias parameter} \label{sec:bhm}

In this section, we compare the mass-weighted halo bias parameter predicted by various theoretical models, all computed using Equation~\ref{eq:theory_bhm}, where several halo bias prescriptions from the \texttt{colossus} package are adopted. We first examine how $b_{\rm hm}$ behaves under different lower limits on group mass, as shown in the left panel of Figure~\ref{fig:biasf_model}. The sample used to estimate $b_{\rm hm}$ is our DESI galaxy group mock catalog in the redshift range $0.0 < z < 0.4$, where we fix the redshift to 0.29 when computing the halo bias. While all models predict a similarly increasing trend of $b_{\rm hm}$ with $M_{\rm th}$, they differ significantly at any given $M_{\rm th}$. This strengthens our motivation to determine the halo bias directly from the field density, rather than relying on a particular theoretical model. We further assess the behavior of $b_{\rm hm}$ when the halo bias is evaluated at different redshifts, for a group sample with $M_{\rm th} > 10^{12.5} \Msunh$ in the range $0 < z < 0.4$. The model-to-model discrepancies at different redshifts further support the conclusion that the computation of $b_{\rm hm}$ from theoretical prescriptions is inherently uncertain.

% In the calculation, we restrict the phase difference from $-\pi$ to $\pi$.

\section{Direction of velocity vector} \label{sec:dir_vel}

In this section, we quantify how well the direction of the reconstructed velocity vector matches the true one by using the cosine of the angle between them, $\mathrm{cos}\ \theta$, defined as

\begin{figure}
    \centering
    \includegraphics[width=0.48\textwidth]{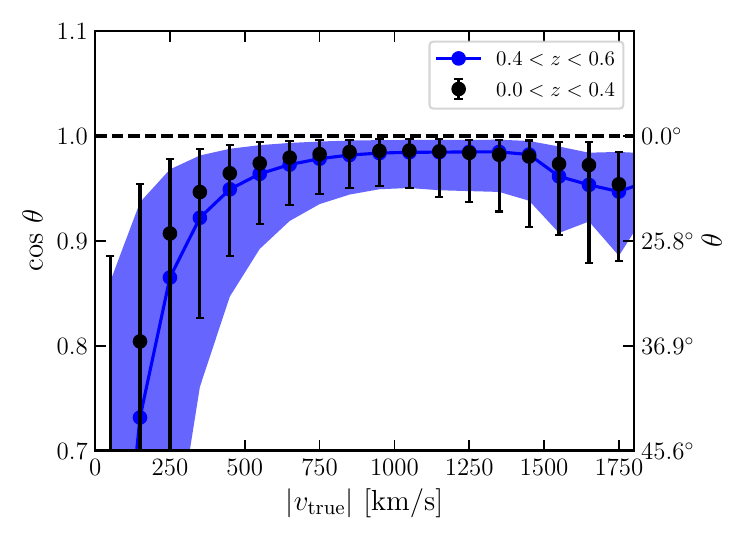}
    \caption{Direction difference between the reconstructed and true velocity vectors in grid cells as a function of the true velocity magnitude. The left and right $y$-axes show the cosine of angle and the corresponding angle between the reconstructed and true velocity vectors, respectively. The black and blue points show the medians at $0.0 < z < 0.4$ and $0.4 < z < 0.6$, with error bar and shadow regions representing corresponding 16th and 84th percentile ranges.}
    \label{fig:vel_angle}
\end{figure}

\begin{flushleft}
\begin{equation}
    \mathrm{cos}\ \theta = \frac{{\bm{v}_{\rm rec}} \cdot \bm{v}_{\rm true}}{|\bm{v}_{\rm rec}||\bm{v}_{\rm true}|}\,,
\end{equation}
\end{flushleft}
 
where $|\bm{v}_{\rm rec}|$ and $|\bm{v}_{\rm true}|$ denote the magnitudes of the reconstructed and true velocity vectors, respectively. As shown in Figure~\ref{fig:vel_angle}, $\mathrm{cos}\ \theta$ stays very close to unity for $|v_{\rm true}| \gtrsim 300 \kms$, implying that the reconstructed velocity vectors are highly aligned with the true ones, even at high velocities ($\sim1500 \kms$), and this holds within both redshift ranges. In the low-velocity regime, where the description of galaxy groups is incomplete because of observational limitations, the median value of $\mathrm{cos}\ \theta$ still approaches about 0.8.

\FloatBarrier

%% For this sample we use BibTeX plus aasjournalv7.bst to generate the

\bibliography{paper}{}
\bibliographystyle{aasjournalv7}



\end{document}